\documentclass[aps,prb,superscriptaddress,twocolumn,showpacs,notitlepage]{revtex4-1}

\usepackage{amsmath}
\usepackage{amsfonts}
\usepackage{amssymb}
\usepackage{hyperref} 
\usepackage{graphicx} 
\usepackage{color}
\usepackage{xcolor}

\usepackage[mathscr]{euscript}%\usepackage[uft8]{inputenc}
\usepackage{rotating}

\newcommand{\ud}[1]{{#1^{\dagger}}}
\newcommand{\bra}[1]{\left\langle #1\right|}
\newcommand{\ket}[1]{\left| #1\right\rangle}

\newcommand\Tr{\mathrm{Tr}}
\newcommand{\mean}[1]{\langle#1\rangle}

\begin{document}

\flushbottom 
%\title{Exciting with Quantum Light. I.~Exciting an oscillator.}
\title{Exciting with Quantum Light. II.~Exciting a two-level system.}
% \title{Exciting with Quantum Light. III.~Exciting with a bundler.}
% \title{Exciting with Quantum Light. IV.~Exciting polaritons.} 
% \title{Exciting with Quantum Light. V.~Pulsed excitations.}

\author{J.~C.~L\'{o}pez~Carre\~{n}o} 
\affiliation{Departamento
de F\'isica Te\'orica de la Materia Condensada, Universidad
Aut\'onoma de Madrid, 28049 Madrid, Spain}

\author{C.~S\'{a}nchez~Mu\~{n}oz} 
\affiliation{Departamento de
F\'isica Te\'orica de la Materia Condensada, Universidad
Aut\'onoma de Madrid, 28049 Madrid, Spain} 

\author{E.~del~Valle}
\affiliation{Departamento de F\'isica Te\'orica de la Materia
Condensada, Universidad Aut\'onoma de Madrid, 28049 Madrid,
Spain} 

\author{F.~P.~Laussy} 
\affiliation{Russian Quantum Center, Novaya 100, 143025 Skolkovo, Moscow Region, Russia}
\affiliation{Departamento de F\'isica Te\'orica de la Materia Condensada, Universidad Aut\'onoma de Madrid, 28049 Madrid, Spain} 

\begin{abstract} 
  We study the excitation of a two-level system (2LS) by quantum
  light, thereby bringing our previous studies~(see part~I. of this
  series) to a target that is quantum itself. While there is no gain
  for the quantum state of the target as compared to driving it with
  classical light, its dynamical features, such as antibunching, can
  be improved.  We propose a chain of two-level systems, i.e., setting
  the emission of each 2LS as the driving source of the following one,
  as an arrangement to provide better single-photon sources. At a
  fundamental level, we discuss the notion of strong-coupling between
  quantum light from a source and its target, and the several versions
  of the Mollow triplet that follow from various types of driving
  light. We discuss the Heitler effect of antibunched photons from the
  scattered light off a laser.
\end{abstract}

\date{\today} 
\maketitle 

%\begin{widetext}
%  \tableofcontents
%\end{widetext}

\section{Introduction} 

The two-level system (also known as the ``two-state quantum
system''~\cite{feynman_book71c}) is an important type of oscillator in
the quantum universe.  In some respects, it can be seen as the truly
quantum oscillator, with a response at the single-particle level,
whereas the (quantum) harmonic oscillator appears instead as a
classical oscillator able to sustain non-classical motion.  The 2LS, a
fundamental object in any case, rose to gigantic proportion with
Deutsch's insight of quantum computation,\cite{deutsch85a} that
turned it into a ``qubit'', the elementary piece of information in the
physical universe. It then became of primary importance to control the
dynamics of a two-level system. To this day, this is achieved mainly
with classical laser pulses, thanks to a ``pulse-area'' theorem that
states that any desired final state can be obtained with a suitable
pulse.\cite{golubev14a}  Entire textbooks have been written on the
topic~\cite{allen_book87a} and still did not exhaust it.

In this text---second in a series where we address the general problem
of exciting with quantum light\cite{arXiv_lopezcarreno16a}---we study
the excitation of the 2LS (or qubit) when driven quantum mechanically,
rather than classically. We focus in the present text on cw excitation
and refer to part~V for pulsed excitation, of particular relevance for
state preparation and quantum information processing. We will show
that while the quantum states that can be prepared in a 2LS do not
benefit from a quantum driving as compared to classical excitation,
the dynamical emission on the other hand can reach new regimes. As an
application, we will show how this can be used to engineer better
single-photon sources. To a large extent, the problem posed in this
text also addresses the fundamental question of what defines
strong-coupling. The excitation of a 2LS by light falls largely in the
framework of resonance fluorescence, whose most notable manifestation
is the so-called Mollow triplet.\cite{mollow69a}  This occurs when
the intense excitation (by a classical laser) of a 2LS dresses its
bare states (ground $\ket{g}$ and excited $\ket{e}$) to give rise to
new eigenstates from their quantum superposition
$\ket{\pm}\equiv(\ket{g}\pm\ket{e}/2$. The coupling is mediated by the
classical laser that, being a $c$-number parameter, does not directly
appear in the structure of the Hilbert space and thus neither in the
quantum state. We will discuss in detail the nature of these dressed
states when quantizing the excitation field and how various quantum
states leads to various results.

The text is organized as follows.
Section~\ref{sec:juefeb25193818CET2016} briefly reminds the formalism
which has been amply motivated and introduced in part~I (Sections~I
and~II)~\cite{arXiv_lopezcarreno16a} and we refer to this text and
references therein for further details.  In
Section~\ref{sec:juefeb25194429CET2016}, we study which states of the
2LS are accessible under quantum excitation where we show how, in the
cw regime, classical excitation is more suitable than driving by, say,
a Single-Photon Source (SPS).  In
Section~\ref{sec:juefeb25195350CET2016}, we discuss how
strong-coupling takes place with the more general case of quantum
light as the driving agent. Beyond a new definition for
strong-coupling needed to accommodate quantization of the driving
light-field, this discussion will also allows us to pinpoint which
features are specific to the classical driving by contrasting all the
variants of the Mollow triplet under various types of excitation. For
instance, we will shed light on the mysterious Heitler effect whereby
coherent absorption-emission processes in the low driving limit result
in photons with the first-order coherence of the source (the laser)
and the second-order coherence of the target (the 2LS). As such a
source of spectrally narrow antibunched photons is of high
technological interest, the understanding of its underlying principle
is important. In Section~\ref{sec:juefeb25200943CET2016}, we take
advantage of this understanding to design better SPS by turning the
target into a source of its own, that excites still another 2LS. We
will show how better antibunching can be achieved in this way that
overcomes the Rayleigh scattering.
Section~\ref{sec:jueoct29112554CET2015} concludes.
\begin{figure}
  \includegraphics[width=0.95\linewidth]{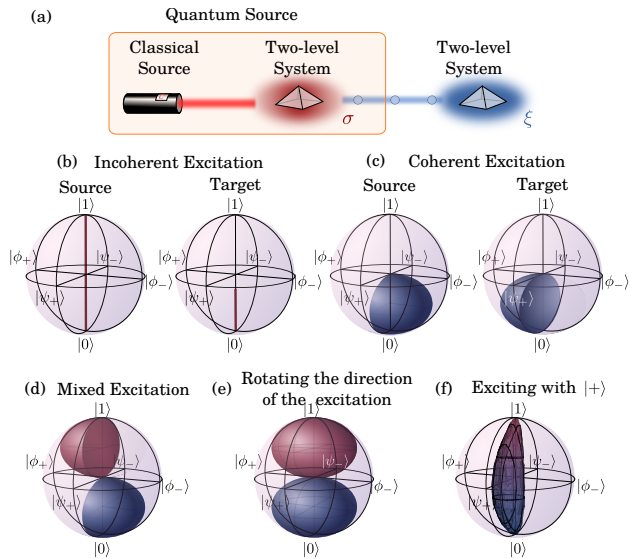}
  \caption{(Color online). (a)~A 2LS is driven by the emission of a
    quantum source. In this particular case, the quantum source is
    made of a 2LS driven by a classical source. (b-f)~Accessible
    states of the 2LS under different driving configurations, as seen
    by the accessible volume in the Bloch sphere.  The states of the
    equator of the sphere are $\ket{\psi_{\pm}} =
    \left(\ket{0}\pm\ket{1}\right)/\sqrt{2}$ and $\ket{\phi_{\pm}} =
    \left(\ket{0}\pm i\ket{1}\right)/\sqrt{2}$.  (b)~Incoherent
    excitation: this kind of excitation does not generate coherence in
    the 2LS, so all the accessible states lie along the $z$--axis of
    the Bloch sphere. (c)~Coherent excitation: this kind of excitation
    does not allow the qubit to have a population larger than $1/2$,
    so all the accessible states lie in the south hemisphere of the
    Bloch sphere. The source 2LS is allowed to go only in one half an
    the ellipsoid, while the target 2LS can cover the ellipsoid
    completely.  (d)~Mixed excitation: under this driving the source
    2LS can access a symmetrical region in the north hemisphere of the
    Bloch sphere.  However, the region available to the target 2LS
    does not increase.  (e)~Rotating the direction of the excitation
    of the source 2LS allows it to access the regions otherwise out of
    reach, thus filling the entire ellipsoid. (f)~Exciting with a
    source that generates the transition $\ket{+}\bra{0}$ allows the
    source 2LS to access states with mean population $1/2$ but not
    completely mixed. }
  \label{fig:fig1}
\end{figure}

\section{Theoretical description}
\label{sec:juefeb25193818CET2016}

Like in part~I,\cite{arXiv_lopezcarreno16a} we use the cascaded
formalism to describe the irreversible excitation of a target by a
quantum source. Here, we focus on a 2LS as the target (with
annihilation operator~$\xi$) and consider various sources that will be
introduced as they are referred to. Taking for instance another 2LS
with annihilation operator~$\sigma$ as the source, as sketched in
Fig.~\ref{fig:fig1}(a), the resulting master equation
takes the form (we take $\hbar=1$ along the paper):
\begin{multline}
\label{Eq.MasterEquation}
\partial_t\rho=i[\rho, H_\sigma + H_\xi] + \frac{\gamma_\sigma}{2}
 \mathcal{L}_\sigma \rho 
+ \frac{\gamma_\xi}{2} \mathcal{L}_\xi \rho +  {}\\
{}+\sqrt{\gamma_\sigma \gamma_\xi} \left \lbrace [\sigma \rho, \ud{\xi}] +
 [ \xi,\rho \ud{\sigma}] \right \rbrace\, .
\end{multline}
Here
$\mathcal{L}_c \rho = (2c\rho\ud{c} - \rho \ud{c}c - \ud{c}c\rho)$,
where $c=\sigma$ for the source and $c=\xi$ for the target, with
$\sigma$, $\xi$ following the Pauli algebra. Each system has a decay
rate and a Hamiltonian, labeled~$\gamma_c$ and~$H_c$. The source must
also be excited, and the details and considerations required to
describe properly the driving of the source are presented in
Section~II of part~I.~\cite{arXiv_lopezcarreno16a} We now proceed to
investigate which regions of the Bloch sphere are accessible under
various types of driving.

\section{Accessible states}
\label{sec:juefeb25194429CET2016}

While we had to introduce in part~I of the series a new charting of
the Harmonic oscillator's Hilbert space to map the states created by
quantum excitation, there has long been a comprehensive representation
for the two-dimensional Hilbert space of the qubit: the \emph{Bloch
  sphere}, that allows to map unambiguously pure states since the
sphere is the projective space for two complex
lines\cite{bengtsson_book08a}~$(\alpha,\beta)\in\mathbf{C}^2$ and
pure states of the 2LS are given by the wavefunction:
\begin{equation}
  \label{eq:sábene16130745CET2016}
  \ket{\psi}=\alpha\ket{0}+\beta\ket{1}\,.
\end{equation}
The mapping is complete since it also accounts for the phase between
the probability amplitudes. The quantum dynamics of a 2LS also evolves
naturally in this geometry,\cite{feynman57a} also accommodating mixed
states \emph{inside} the Bloch sphere (pure states lie on its
surface).  Mixed states are expressed by a density matrix, which is
also determined by two numbers: the population and the coherence of
the qubit:
\begin{equation}
  \rho = \left ( 
    \begin{array}{cc}
      1-n_\sigma & \mean{\sigma}^{\ast}  \\
      \mean{\sigma} & n_\sigma \end{array}\right)\, ,
  \label{Eq.DensityMatrix}
\end{equation}
where $n_\sigma\equiv\mean{\ud{\sigma}\sigma}$, the 2LS population, is
such that $0\le n_\sigma\leq 1 $, and $\mean{\sigma}$, the coherence
of the qubit, is a complex number.  Since the eigenvalues of the
density matrix must be non--negative, the coherence of the qubit has
an upper bound: $|\mean{\sigma}|^2 \leq n_\sigma(1-n_\sigma)$. The
state in Eq.~(\ref{Eq.DensityMatrix}) is pure if $\Tr({\rho^2})=1$,
i.e., if $|\mean{\sigma}|^2 = n_\sigma(1-n_\sigma)$, in which case the
states in Eq.~(\ref{eq:sábene16130745CET2016}) and
Eq.~(\ref{Eq.DensityMatrix}) are equivalent, with $n_\sigma=|\beta|^2$
and $\mean{\sigma} = \alpha \beta^{\ast}$. Thus, any state in the form
of Eq.~(\ref{Eq.DensityMatrix}) is represented in the Bloch sphere as
a point with cartesian coordinates:
\begin{equation}
  \label{eq:lunmay16190238CEST2016}
  x=2\mathcal{R}\mean{\sigma}\,\quad
  y=-2\mathcal{I}\mean{\sigma}\quad\text{and}\quad
  z=1-2n_\sigma\,,
\end{equation}
%
% \begin{subequations}
%   \begin{align}
%     x&=2\mathcal{R}\mean{\sigma}\, ,\\
%     y&=2\mathcal{I}\mean{\sigma}\, ,\\
%     z&=1-2n_\sigma\, ,
%   \end{align}
% \end{subequations}
such that $x^2+y^2+z^2\leq 1$. The amount of vacuum and excitation,
often referred to as \textit{population imbalance}, varies along the
$z$-axis. The relative phase between the vacuum and the excited state
is given by $\phi =\arg{\mean{\sigma}}$ and varies around the
around~the~$x$--$y$~plane.

\subsection{Incoherent driving}

To contrast the excitation of a 2LS by quantum light with the usual
counterpart of classical excitation, we must first remind (and in some
cases possibly derive for the first time) the situation in the latter
case. We will see throughout that the need to understand the quantum
excitation also teaches us on features of the more familiar classical
case.

The expected value of the coherence, $\mean{\sigma}$, of an
incoherently driven 2LS is zero, so its steady--state density matrix
is fully determined by its population only:
\begin{equation}
\label{Eq.PopulationIncoherent}
n_\sigma^{\mathrm{inc}} = \frac{P_\sigma}{P_\sigma+\gamma_\sigma}\, .
\end{equation}

The accessible states under incoherent excitation cover the entire
$z$--axis, as shown in the Bloch sphere at the left of
Fig.~\ref{fig:fig1}(b). Using this 2LS as a source for another 2LS
reduces the span of accessible area. Namely, the steady--state of the
target 2LS, also determined only by its population, reads:
 \begin{equation}
   n_\xi^{\mathrm{inc}} = \frac{4n_\sigma^{\mathrm{inc}} \gamma_\sigma(P_\sigma+\gamma_\sigma+\gamma_\xi)}{P_\sigma^2+(\gamma_\sigma+\gamma_\xi)^2+4\Delta_{\sigma\xi}^2+2P_\sigma(\gamma_\xi
+5\gamma_\sigma)}\, ,
\end{equation}
where $\gamma_\xi$ is the decay rate of the target 2LS, and
$\Delta_{\sigma\xi} = \omega_\sigma - \omega_\xi$ is the detuning
between the source and the target.  The accessible states are shown in
the Bloch sphere at the right Fig.~\ref{fig:fig1}(b), and unlike the
source 2LS that can be saturated to its excited state by incoherent
pumping, the population of the target 2LS lies between $0\leq
n_\xi^{\mathrm{inc}} \leq 0.3535$. The upper bound $0.3535\approx 7/20$
is obtained in the regime $P_\sigma \approx \gamma_\sigma$ for the
source, in which case $n_\sigma^{\mathrm{inc}}=1/2$, and when
$\gamma_\xi\approx\gamma_\sigma$. The reason for this saturation is a
type of self-quenching, here induced by the power broadening of the
source. This shows clearly the intrinsic limitations of exciting with
a 2LS. One can increase the emission rate, but this comes at the
expense of other fundamental parameters such as the spectral
broadening. In contrast, one can increase the emission rate of a
cavity independently of its spectral width.

\subsection{Coherent driving}
 
In the case of coherent classical driving of the source, its
population and coherence are given by:
\begin{subequations}
  \begin{align}
    n_\sigma^{\mathrm{coh}}&= \frac{4\epsilon_1\Omega^2}{ \gamma_\sigma^2+4\widetilde\omega_\sigma^2+8\epsilon_1\Omega^2}\, ,\\
    \mean{\sigma}^{\mathrm{coh}}&=\frac{2\sqrt{\epsilon_1}
\Omega(2\widetilde\omega_\sigma+
      i \gamma_\sigma)}{\gamma_\sigma^2+4\widetilde
      \omega_\sigma^2+8\epsilon_1\Omega^2}\, ,
\end{align}
\label{Eq.PopulationCoherent}
\end{subequations}
where $\Omega$ is the rate of coherent excitation, $\epsilon_1$ is
the amplitude of the channel through which we drive coherently the
source, and $\widetilde\omega_\sigma$ is the detuning between the
driving laser and the 2LS.  Unlike the incoherent excitation, these
states now span a volume in the Bloch sphere, which means that they
have some degree of purity, the higher the closer to the ground state.
Also, since coherent driving forbids population inversion, this volume
is restricted to the south hemisphere of the Bloch sphere, reaching
population $n_\sigma=1/2$ at most in a maximally-mixed state. Finally,
the phase of the driving laser also fixes the phase of the excitation
of the 2LS, so only one half of the southern hemisphere is
accessible. The final accessible volume is shown in the left Bloch
sphere in Fig.~\ref{fig:fig1}(c). Its boundaries are given by the
shell of an ellipsoid with equation:
\begin{equation}
  1= \left(\frac{x}{\sqrt{2}/2}\right)^2+\left(\frac{y}{\sqrt{2}/2}\right)^2
  +\left(\frac{z-z_0}{1/2}\right)^2\, ,
  \label{Eq.LimitsCoherent}
\end{equation}
with $z_0=-1/2$, $|x| \leq \sqrt{2}/2$ and $-\sqrt{2}/2\leq y
\leq 0$.
 
The expressions for the population and coherence of the target 2LS are
not as straightforward as for the source 2LS, and although they can be
found in closed form, they are too lengthy to be written here. As an
illustration of their complexity, we provide the particular case when
the driving laser, the source 2LS, and the target 2LS are in
resonance, in which case the population and the coherence of the
target 2LS reduce to:
\begin{widetext}
  \begin{subequations}
    \label{eq:FriSep2234841CEST2016}
    \begin{align}
      n_\xi^{\mathrm{coh}}&= 16\epsilon_1\Omega^2(1-\epsilon_1)\gamma_{\widetilde{10}} \left[
        \gamma_{\widetilde{11}}^3 \gamma_{\widetilde{21}}\gamma_{\widetilde{12}}+
        8\epsilon_1\Omega^2(2\gamma_{\widetilde{10}}^3+6\gamma_{\widetilde{10}}^2\gamma_{\widetilde{01}}+
\gamma_{\widetilde{10}}\gamma_{\widetilde{01}}^2(1-\epsilon_1)+3\gamma_{\widetilde{01}}^3)+64\gamma_{\widetilde{01}}\epsilon_1^2\Omega^4\right ] 
\big / N^\ast\, , \\ 
\mean{\xi}^{\mathrm{coh}} &=
      -4i\sqrt{\gamma_{\widetilde{10}}\gamma_{\widetilde{01}} \epsilon_1(1-\epsilon_1)}\Omega\gamma_{\widetilde{10}} \left\lbrace
        \gamma_{\widetilde{11}}^3\gamma_{\widetilde{12}}\gamma_{\widetilde{21}}+8\gamma_{\widetilde{11}}^2\epsilon_1\Omega^2\left[4\gamma_{\widetilde{01}}+\gamma_{\widetilde{10}}(8\epsilon_1-3)\right]
        +128\epsilon_1^2\Omega^4\left[\gamma_{\widetilde{01}}+\gamma_{\widetilde{10}}(2\epsilon_1-1) \right] \right \rbrace \big / N^\ast\, ,\\ 
\mathrm{with} \quad N^\ast&=\gamma_{\widetilde{01}}\gamma_{\widetilde{10}}^2\gamma_{\widetilde{11}}^3\gamma_{\widetilde{21}}\gamma_{\widetilde{12}}+8\gamma_{\widetilde{11}}^2
\epsilon_1\Omega^2\left[2\gamma_{\widetilde{01}}^4+7\gamma_{\widetilde{01}}^3\gamma_{\widetilde{10}}+12\gamma_{\widetilde{01}}^2
\gamma_{\widetilde{10}}^2+2\gamma_{\widetilde{01}}\gamma_{\widetilde{10}}^3(13-10\epsilon_1) +8\gamma_{\widetilde{10}}^4
(1-\epsilon_1)\right]+64 \epsilon_1^2\Omega^4\left[5\gamma_{\widetilde{01}}^4+{}\right. 
\nonumber\\
&{}+\left. 2\gamma_{\widetilde{01}}\gamma_{\widetilde{10}}^3
(23-20\epsilon_1)+3\gamma_{\widetilde{01}}^2\gamma_{\widetilde{10}}^2(21-16\epsilon_1)+2\gamma_{\widetilde{01}}^3
\gamma_{\widetilde{10}}(17-10\epsilon_1)+8\gamma_{\widetilde{10}}^4\right] + 1024 \epsilon_1^3\Omega^6
\left[\gamma_{\widetilde{01}}^2+\gamma_{\widetilde{01}}\gamma_{\widetilde{10}}(3-2\epsilon_1)\right]\, ,
    \end{align}
  \end{subequations}
\end{widetext}
where we have introduced the notation:
\begin{equation}
  \label{eq:juejun16145638CEST2016}
  \gamma_{\widetilde{mn}}^k \equiv (m\gamma_\sigma+n \gamma_\xi)^k\,,
\end{equation}
e.g., $\gamma_{\widetilde{21}}^2 = (2\gamma_\sigma +
\gamma_\xi)^2$. Interestingly, while
Eqs.~(\ref{eq:FriSep2234841CEST2016}) for the target have a much more
complicated form than Eqs.~(\ref{Eq.PopulationCoherent}) for the
source, they lead to the same accessible volume of the type of
Eq.~(\ref{Eq.LimitsCoherent}). At such, unlike the case of incoherent
excitation, the target suffers no restriction of its accessible region
under quantum excitation by a coherent excited 2LS as compared to
direct excitation from the coherent source. In fact, since the phase
of excitation is not fixed by the source, the accessible volume of the
target 2LS is given by the full ellipsoid in
Eq.~(\ref{Eq.LimitsCoherent}) with $|y|\leq \sqrt{2}/2$, as shown in
the right Bloch sphere of Fig.~\ref{fig:fig1}(c).

\subsection{Mixture of coherent and incoherent driving}

The source 2LS can drive its target from the entire $z$--axis when
driven incoherently to half an ellipsoid in the southern hemisphere
when driven coherently. Mixing the two types of excitations allows the
source 2LS to access a volume in the northern hemisphere as well,
symmetric to the volume accessible by coherent excitation only, as
shown in Fig.~\ref{fig:fig1}(d). The expressions for the population
and the coherence in this case are:
\begin{subequations}
\begin{align}
n_\sigma^{\mathrm{mix}}&= \frac{P[(P+\gamma_\sigma)^2+4\widetilde\omega_\sigma^2]+
4\epsilon_1\Omega^2(P+\gamma_\sigma)}{(P+\gamma_\sigma)[(P+\gamma_\sigma)^2+
4\widetilde\omega_\sigma^2+8\epsilon_1\Omega^2]}\, ,\\
\mean{\sigma}^{\mathrm{mix}}&=\frac{2\sqrt{\epsilon_1}\Omega(P-\gamma_\sigma)[2\widetilde\omega_\sigma+i(P+\gamma_\sigma)]}{(P+\gamma_\sigma)[(P+\gamma_\sigma)^2+4\widetilde\omega_\sigma^2+8\epsilon_1\Omega^2]}\, ,
\end{align}
\end{subequations}
from which we see that the condition to access the volume in the
northern hemisphere, i.e., $n_\sigma \geq 1/2$, is simply that
$P_\sigma>\gamma_\sigma$.  The northern-hemisphere volume is enclosed
by the ellipsoid in Eq.~(\ref{Eq.LimitsCoherent}) with $z_0=1/2$,
$|x| \leq \sqrt{2}/2$ and $0\leq y \leq \sqrt{2}/2$.

Using this 2LS under joint coherent and incoherent driving to bring
its steady state off-axis in the northern hemisphere does not,
however, make it a quantum source that can drive a target 2LS beyond
the southern-hemisphere ellipsoid of coherent excitation,
Fig.~\ref{fig:fig1}(c). The expressions for the population and
coherence of the target 2LS in this case are analytical as well, but
even at resonance and using the compact notation of
Eq.~(\ref{eq:juejun16145638CEST2016}), they are too bulky to be
written here.  Note that by changing the direction from which the
laser drives the source 2LS, or by using waveplates, one can rotate
its phase, changing $\sigma$ for $e^{i \theta} \sigma$ in the master
equation~(\ref{Eq.MasterEquation}). This variation leaves the
population of the source 2LS unchanged, but adds a phase to its
coherence. In this way, the source 2LS can span all states in a volume
obtained by revolving around the $z$--axis.  The same holds for the
case of mixed excitation, and therefore the total accessible volume of
the source 2LS is given by two full ellipsoids as shown in
Fig.~\ref{fig:fig1}(e).

Overall, these results show that a 2LS as a source of excitation is no
more advantageous than classical sources as far as accessible volumes
in the Bloch sphere and in the steady state are concerned. The target
is ``quantum-enough'' not to benefit further from quantum excitation
(at least when it comes from a SPS) as compared to classical
excitation. We will see later in the text that this is because we
consider the target in isolation, and that including quantum
correlations with the source indeed lead to departures and benefits
from a quantum driving.  In the next paragraph, we also anticipate on
more general results to be presented in part~IV of this Series that
considers the excitation from more exotic quantum sources than simply
a SPS, in which case, the target's quantum state can depart from the
classical driving cases.

\subsection{Phenomenological quantum source}

So far we have modeled the quantum source by describing it fully and
self-consistently, coupling its output to a target through the
cascaded formalism. The excitation of the source itself was achieved
phenomenologically from the Lindblad formalism. One could also in
principle take a simpler route by describing directly the quantum
source through ad-hoc Lindblad terms. This is an expedient way to see
whether quantum excitation can bring us to regions beyond those of
access from the classical case, that lie in an ellipsoid of, for most
of the cases, mixed states (not on the surface of the sphere). In
particular, a common feature of all the kinds of excitation we have
considered is that the state with population $1/2$ is completely
mixed, i.e., it is located at the origin of the sphere. 

We now show how to drive the target 2LS to reach states with
population $1/2$ but with nonzero coherence. Specifically, we assume a
source that emits photons in the state $\ket{+}_\phi = (\ket{0} + e^{i
  \phi}\ket{1})/\sqrt{2}$. At the simplest level, the incoherent type
of such an excitation is described with Lindblad terms in the master
equation~(\ref{Eq.MasterEquation}) of the type $(P^{\ast}/2)
\mathcal{L}_x \rho$, where $x = \ket{+}_\phi \bra{0}$ is the operator
that brings the vacuum state into $\ket{+}_\phi$, and $P^{\ast}$ is
the rate at which this excitation is enforced onto the target. The
resulting population and coherence of the source 2LS in this case are
given by:
\begin{subequations}
  \begin{align}
    n_\sigma^{+}&= \frac{P^{\ast}(P^{\ast}+
      2\gamma_\sigma e^{i \phi})}{P^{\ast\, 2} + 4\gamma_\sigma^2
      +4 P^{\ast}\gamma_\sigma \cos 2\phi}\, ,\\
    \mean{\sigma}^{+}&=\frac{2P^\ast\gamma_\sigma}
    {(P^\ast+\gamma_\sigma)(P^\ast e^{2i \phi}+2\gamma_\sigma)}\, ,
  \end{align}
\end{subequations}
where the phase $\phi$ must be chosen so that $n_\sigma^+ \in
\mathbb{R}$.  The accessible volume with $\phi=0$ is shown in
Fig.~\ref{fig:fig1}(f). This case demonstrates a type of quantum
excitation that drives the 2LS target into a volume of the Bloch
sphere that strongly differs from the classical counterpart.  The
cw-excitation spoils the coherence, and even a source that explicitly
drives the 2LS into the $\ket{+}$ state cannot sustain completely its
coherence of the 2LS. Nevertheless, this indicates that the general
case of quantum excitation does ultimately bring us further than the
classical case, even though the SPS does not. Importantly, to clarify
this point at this stage, we have assumed a simple phenomenological
model for a particular case.  It is not clear how such a source could
be devised from an Hamiltonian in the first place before it is plugged
to its target through the cascaded formalism. For instance, a close
counterpart that emits photons in the state $\ket{+}_\phi = (\ket{0} +
e^{i \phi}\ket{1})/\sqrt{2}$ with a complex superposition
($\phi\not\in\{0,\pi\}$), that is easily conceived conceptually,
cannot be described in the above Lindblad form since this results in
non-real coefficients of the master equation that yield unphysical
density matrix (and, e.g., complex populations).

\section{Mollow dressing}
\label{sec:juefeb25195350CET2016}

\begin{table*}[t]
  \begin{ruledtabular}
    \begin{tabular}{l|l|l}
      Source & Target & Treated in \\
      \hline
      Classical laser & Mollow triplet & Section~\ref{sec:viemay13191527CEST2016} \\
      Two-level system excited incoherently & Unresolved Mollow triplet & Section~\ref{sec:viemay13192349CEST2016} \\
      Two-level system excited coherently & Mollow triplet with attenuated sidebands & Section~\ref{sec:viemay13192407CEST2016} \\
      Cavity spontaneous emission & Variant of the Mollow triplet & Op.~V \\
      Cavity excited incoherently & Deformed singlet & Section~\ref{sec:viemay13191706CEST2016} \\
      Cavity excited coherently & Mollow triplet & Section~\ref{sec:TueAug30175459CEST2016} \\
%      Cavity excited incoherently & Mollow & here \\      
      One-atom laser & Quintuplet to Mollow triplet transition & Section~\ref{sec:viemay13191827CEST2016} \\
      $N$-photon source & Quantum Mollow triplet & Op.~IV \\
    \end{tabular}
  \end{ruledtabular}
  \caption{Classification of the shape of the emission of the target
    2LS according to the source driving it.}
  \label{tab:mollows}
\end{table*}

\subsection{Introduction}

In this Section, we go beyond which average quantum state a target 2LS
can be driven into to consider instead some deeper structural aspect,
namely, we describe the energy spectrum and its associated set of
states. The naked 2LS has the simplest possible structure of a
vacuum~$\ket{\mathrm{g}}$ and an excited
state~$\ket{\mathrm{e}}$. When placed inside a cavity, in absence of
dissipation, a new set of quantum states for the combined system takes
over, that are eigenvectors for the coupling
Hamiltonian:\cite{jaynes63a}
$(\ket{\mathrm{g}n}+\ket{\mathrm{e},n-1})/\sqrt2$ at resonance
with~$n$ the number of photons in the cavity. The respective pair of
energies constitutes one stair of the so-called Jaynes--Cummings
ladder, that repeats this structure for each integer~$n$. These states
are entangled and go by the name of ``polaritons'' or ``dressed
states''. In presence of dissipation, a notion of weak-coupling and
strong-coupling emerges to distinguish cases where bare states
$\ket{\mathrm{g}, \mathrm{e}}$ and $\ket{n}$ dominate the dynamics as
dissipation destroyed their quantum superpositions. In the former case
the coupling is classical while in the latter it binds the states
through quantum nonlocality.  This is a central concept of cavity QED
but one that remains vaguely defined as rooted in the simple
Hamiltonian framework. When including dissipation, the dressed-states
energies become complex as a result of their finite lifetime, and
acquire a broadening specific to their constitution in terms of bare
states in addition to a renormalized energy. This is well understood
for two coupled oscillators but already the textbook case of the 2LS
in a cavity displays a much less familiar structure, the so-called
``dissipative Jaynes--Cummings ladder'',\cite{delvalle09a} with
coexistence of weak and strong-coupling depending on the manifold of
excitation.  Such a description can be applied to other
systems.\cite{quesada12a} When including, beyond mere decay, also an
excitation scheme, the situation can become extremely
complex.\cite{delvalle11a} The simplest problem of this type is the
celebrated Mollow triplet,\cite{mollow69a} where a 2LS is strongly
excited by a classical field ($c$-number). This is the context in
which the term of ``dressed state'' appeared (from
Cohen-Tannoudji~\cite{cohentannoudji77a}). In this text where we study
quantum driving of a 2LS, we are naturally brought to consider how the
quantum features of light affect the 2LS target's state. In the limit
where light is coherent, we expect to recover the conventional Mollow
scenario, as is already known to be the case from the Jaynes--Cummings
perspective where light is quantized.\cite{delvalle10d}  

We will now turn to the more general problem of how a 2LS target
becomes affected by quantum excitation without feedback, i.e., beyond
the Hamiltonian formalism (the cases that will be dealt with are
summarized in Table~\ref{tab:mollows}). This poses some immediate
questions such as: Can a single-photon source dress a 2LS? Are Rabi
oscillations---exchanges of excitation between the modes---necessary
for state-dressing? If so, how can this be achieved in the cascaded
formalism where feedback is precisely forbidden? Does the Mollow
theory of a $c$-number description of the exciting laser break down in
some regime, say of low excitation? These and other fundamental
questions are answered in the remaining of this Section.

\subsection{Complex energy spectrum}

We define strong-coupling as the emergence of resonances in the system
with real energies different from those of the bare states.  Such
resonances are obtained in a dissipative system from the
eigenvalues~$D_p$ of the Liouvillian matrix $M$,\cite{laussy09a} that
follows from writing Eq.~(\ref{Eq.MasterEquation}) as:
\begin{equation}
  \label{eq:M}
  \partial_t\langle \langle \rho \rangle \rangle = -M \langle
  \langle\rho \rangle \rangle\, ,
\end{equation}
where $\langle \langle \rho \rangle \rangle$ is the density matrix laid out in
vectorial form, namely,
$\langle \langle \rho \rangle \rangle=(\rho_{11},\rho_{21},\cdots,\rho_{n1},\rho_{21}\cdots\rho_{nn})^T$
for some truncation~$n$ that can be taken to go to~$\infty$. This
gives access to the ``transitions'' in the system, rather than
directly to the energies of the states. This is a nuance already
present in the first considerations of light-matter
interactions.\cite{weisskopf30a} From the knowledge of the
transitions, however, one can usually reconstruct the underlying
energy structure of the system, that is, both the composition of the
dressed states and their associated energies. A typical observable is
the total emission emission spectrum that results from the combined
emission between the various states:
\begin{equation}
  \label{eq:Sw}
  S(\omega) = \frac{1}{\pi}\sum_p 
  \frac{L_p (\gamma_p/2) - K_p
    (\omega_p - \omega) }{(\gamma_p/2)^2
    +(\omega_p - \omega)^2}\, ,
\end{equation}
where $D_p\equiv \gamma_p/2 + i \omega_p$ is the $p$th resonance,
i.e., eigenvalue of~$M$, that defines the energy structure of the
system, and $L_p$ and~$K_p$ are the corresponding weights for this
transition that determine its prominence in the total emission
accordingly with the state and dynamics of the system. The term~$L_p$
weights a pure Lorentzian emission, corresponding to spontaneous
emission from the initial state towards the final one, while $K_p$
brings a dispersive correction that is typical of coupled oscillators
and that results in our case from interferences between transitions
that overlap in energy. In finite-size Hamiltonian systems or those
that can be decomposed into a direct sum of uncoupled manifolds, such
as the Jaynes--Cummings Hamiltonian, the most important structure
comes from~$D_p$ alone. In infinite-size systems where the system
cannot be closed in a self-consistent way, for instance because an
excitation term connects all the manifolds, one then needs to weight
the transition~$D_p$ with~$|L_p+iK_p|$ as otherwise different
truncation schemes of~$M$ give different results for the complex
energy spectrum, whereas the weighted transitions converge to a
physical result. For this reason, the complex weight is fundamental as
well. It is obtained as:~\cite{laussy09a}
\begin{equation}
  L_p + i K_p=\frac{1}{n_\sigma}\Tr \left 
    \lbrace \sigma \Big[ \Big[ E_{ip}\sum_{n} 
      E^{-1}_{pn} \langle \langle \rho_{ss}\ud{\sigma}\rangle \rangle_n \Big]\Big] \right \rbrace \, ,
\end{equation}
%
% \begin{equation}
%   L_p + i K_p=\frac{1}{n_\sigma}\Tr \left 
%     \lbrace \sigma \Rvzigzag E_{ip}\sum_{n} 
%       E^{-1}_{pn} \rvzigzag\rho_{ss}\ud{\sigma}\lvzigzag_n \Lvzigzag \right \rbrace \, ,
% \end{equation}
%
% \begin{equation}
%   L_p + i K_p=\frac{1}{n_\sigma}\Tr \left 
%     \lbrace \sigma \left\rWavy E_{ip}\sum_{n} 
%     E^{-1}_{pn} \Big\rwavy\rho_{ss}\ud{\sigma}\Big\lwavy_n \right\lWavy \right \rbrace \, ,
% \end{equation}
%
where $\rho_{ss}$ is the steady-state solution of the master equation,
$\sigma$ is the annihilation operator of the 2LS,
$n_\sigma = \Tr(\ud{\sigma}\sigma\rho_{ss})$ is its population and $E$
is the matrix of eigenvectors of~$M$ (i.e., $E^{-1}ME$ is the diagonal
matrix of complex energies). Here
$\langle \langle \mathcal{M}\rangle \rangle_n$ refers to the $n$th
element of the flattened matrix~$\mathcal{M}$ and~$[[ \mathcal{V} ]]$
refers to the reversed process that shapes the vector~$\mathcal{V}$
(of size $n^2$) into a matrix as
$((\mathcal{V}_1, \mathcal{V}_2, \cdots, \mathcal{V}_n),
(\mathcal{V}_{n+1},\mathcal{V}_{n+2},\cdots \mathcal{V}_{2n}),
\cdots)$.

In the following Sections, we start a comprehensive analysis of the
complex energy spectrum, and the structure of states (bare and
dressed) associated to it, for the configurations of excitation of a
2LS listed in Table~\ref{tab:mollows}. Namely, in
Section~\ref{sec:viemay13191527CEST2016} we start with the
conventional Mollow triplet, where a classical field drives a 2LS,
then in Section~\ref{sec:viemay13192349CEST2016} we replace the
classical field by the simplest quantum light, i.e., that emitted by a
SPS; in Section~\ref{sec:viemay13192407CEST2016}, the SPS is brought
itself in the Mollow triplet regime.  Then we come back to the
conventional Mollow configuration of light exciting a 2LS, but with
light described by an operator rather than by a $c$-number. This will
allow us to i)~see what is lost in the approximation of describing the
laser as a sine-wave and ii)~consider classical excitation beyond
merely a coherent state. Specifically, we consider in
Section~\ref{sec:viemay13191706CEST2016} a thermal state of the
light-field (chaotic light), in
Section~\ref{sec:TueAug30175459CEST2016} a coherent state (the closest
to the conventional Mollow case), and, finally, in
Section~\ref{sec:viemay13191827CEST2016}, we take an additional step
in describing the quantum dynamics of an actual laser acting as the
source to drive the 2LS, taking the simplest case of a one-atom laser
to do so. All these cases bring some variations to the problem. Their
common features give a picture of what constitutes the substance of
the Mollow triplet.

\subsection{Excitation by a classical field}
\label{sec:viemay13191527CEST2016}

The simplest classical excitation of a 2LS is that provided by a
thermal source, or incoherent pumping, that brings the system into its
excited state at a rate~$P_\sigma$. For completeness, we address this
case as well before we turn to coherent excitation that leads to
Mollow physics.  The 2LS incoherently pumped is described by the
master equation:
\begin{equation}
\partial_t \rho = i[\rho,\omega_\sigma \ud{\sigma}\sigma] + 
\frac{\gamma_\sigma}{2}\mathcal{L}_{\sigma}\rho+
\frac{P_\sigma}{2}\mathcal{L}_{\ud{\sigma}}\rho\, ,
\label{Eq.MasterEquationIncoherent}
\end{equation}
where $\omega_\sigma$ is the free energy of the 2LS and
$\gamma_\sigma$ its decay rate (inverse lifetime of the excited
state), with $M$-matrix (in Eq.~(\ref{eq:M}):
\begin{equation}
M = \left( \begin{array}{cccc}
P_\sigma & 0 & 0 & -\gamma_\sigma \\
0 & \Gamma_\sigma/2+i\omega_\sigma & 0 &0 \\
0 & 0 & \Gamma_\sigma/2+i\omega_\sigma & 0\\
-P_\sigma & 0 & 0 & \gamma_\sigma
\end{array} \right)\, ,
\end{equation}
where $\Gamma_\sigma\equiv\gamma_\sigma + P_\sigma$.  The result is in
this case trivial, as the only eigenvalue with nonzero weight is
$D_1 = \Gamma_\sigma/2 + i \omega_\sigma$ with weight $L_1 + i K_1 =1$
purely Lorentzian. Note how including the complex weight in the
computation of the complex energy spectrum allows to retain the one
relevant resonance while the matrix~$M$ otherwise features four.  As a
conclusion, a 2LS excited incoherently is simply excited and release
this excitation with the same energy by spontaneous emission, with a
photoluminiscence spectrum given by:
\begin{equation}
  \label{eq:PLIncoherent2LS}
  S(\omega)=
  \frac{1}{\pi}\frac{\Gamma_\sigma/2}{(\Gamma_\sigma/2)^2+
    (\omega\sigma-\omega)^2}\, .
\end{equation}
There is no dynamical effects, shifts, dressing, renormalization of
any sort. We now see how this changes considerably when upgrading the
incoherent excitation to a coherent one.

\begin{figure}
  \includegraphics[width=1\linewidth]{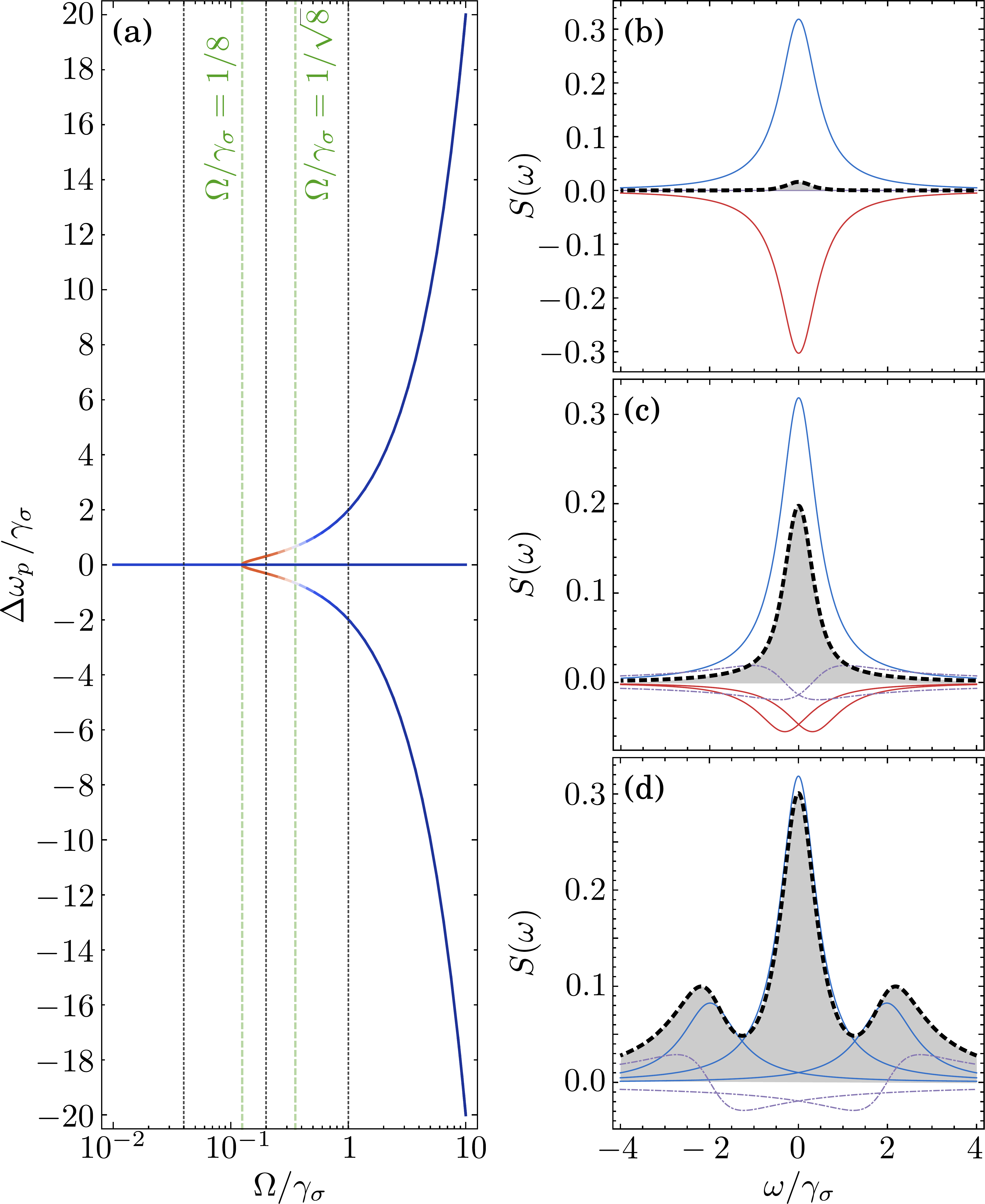}
  \caption{(Color online). Excitation by a coherent source. (a)~Energy
    spsectrum of a coherently driven 2LS. The energies that contribute
    to the total emission spectrum with positive lorentzians are shown
    in blue, whereas those contributing as negative lorentzians are
    shown in red. (b-d) Emission spectra of the target 2LS (dashed,
    black lines) for the values $\Omega/\gamma_\sigma$ marked by the
    vertical dashed lines in Panel~(a). The total emission spectum is
    made of the sum of positive (blue lines) and negative (red)
    lorentzians, and dispersive functions (dotted purple).}
    \label{Fig:fig2}
\end{figure}

The excitation of the 2LS by a coherent classical field, i.e., a
$c$-number~$\Omega\exp(i\omega_\mathrm{L}t)$ with $\Omega^2$ the
intensity of the driving laser, brings us to the Mollow master
equation:
\begin{equation}
  \partial_t \rho = i[\rho, \Omega (\sigma+\ud{\sigma})] +
  \frac{\gamma_\sigma}{2}\mathcal{L}_{\sigma}\rho\,,
  \label{Eq.MasterEquationCoherent}
\end{equation}
in the rotating frame and assuming that the 2LS and the laser are in
resonance. Equation~(\ref{Eq.MasterEquationCoherent}) yields the
following matrix~$M$:
\begin{equation}
  M = \left( \begin{array}{cccc}
      0 & -i \Omega & i \Omega & -\gamma_\sigma \\
      -i \Omega & \gamma_\sigma/2 & 0 &i \Omega \\
      i \Omega & 0 & \gamma_\sigma/2 & -i\Omega\\
      0 & i \Omega & -i\Omega & \gamma_\sigma
    \end{array} \right)\, ,
\end{equation}
which has a more complex set of eigenvalues:
\begin{subequations}
  \label{eq:DMollow}
  \begin{align}
    D_1&=0\,,\\
    D_2&=\gamma_\sigma/2\,,\\
    D_{\pm}&=\frac{1}{4}\left(3\gamma_\sigma\pm \sqrt{\gamma_\sigma^2 -64
      \Omega^2}\right)\,.
  \end{align}
\end{subequations}
The imaginary part of these quantities, that correspond to the
energies of the transitions (real parts correspond to their
broadening), are shown as a function of the intensity of the driving
laser in Fig.~\ref{Fig:fig2}(a). Their corresponding weight is encoded
in the colour: positive weights have blue shade while negative ones
have a red shade.

This simple structure subtends a complex and rich phenomenology.  At
low intensity, $D_{\pm} \approx \gamma_\sigma(3\pm1)/4$ and the
emission spectrum is as shown in Fig.~\ref{Fig:fig2}(b), where $D_2$
is shown as the blue lorentzian and $D_{-}$ is shown as the red
negative lorentzian. The peak due to $D_1$ corresponds to the light
scattered by the 2LS from the laser (not shown) and is known as the
Rayleigh peak. This peak has zero linewidth, and is involved in the
Heitler process that generates antibunched photons with the coherence
of the driving laser.\cite{heitler_book44a}

In this case, and as long as $\Omega \leq \gamma_\sigma/8$, the
spectrum consists of a single line as the eigenvalues's imaginary
parts (that correspond to the states energies) are all degenerate. The
combination of Eqs.~(\ref{eq:Sw}) and (\ref{eq:DMollow}) yields:
\begin{multline}
  S(\omega) =\frac{1}{2\pi}\frac{\gamma_\sigma/2}{(\gamma_\sigma/2)^2+\omega^2}{}\\{}+\frac{1}{4\pi\beta}\frac{(3\gamma_\sigma-\beta)/4}{[(3\gamma_\sigma-\beta)/4]^2+\omega^2}\frac{8(\beta+5\gamma_\sigma)\Omega^2-
  \gamma_\sigma^2(\gamma_\sigma+\beta)}{8\Omega^2+\gamma_\sigma^2}+{}\\
  {}+\frac{1}{4\pi\beta}\frac{(3\gamma_\sigma+\beta)/4}{[(3\gamma_\sigma+\beta)/4]^2+\omega^2}\frac{8(\beta-5\gamma_\sigma)\Omega^2+
  \gamma_\sigma^2(\gamma_\sigma-\beta)}{8\Omega^2+\gamma_\sigma^2}\, ,
\end{multline}
where $\beta=\sqrt{\gamma_\sigma^2-64\Omega^2}$. This total emission
spectrum is shown with the black dashed line in Fig.~\ref{Fig:fig2}(b)
along with its decomposition into the underlying transitions (here not
splitted). The dressed states are largely compensating each
other. This is the process of coherent absorption and re-emission that
endows this regime with peculiar properties. Beside, in this regime
the light emitted by the 2LS is essentially coming from the Rayleigh
peak (not shown). As a result of this structure of emission from the
2LS, the light emitted is antibunched and originates mainly from the
Rayleigh peak, that has the first-order coherence of the laser (it is
a~$\delta$ peak for a $c$-number driving field). This is a strong
feature of resonance fluorescence that has been observed by several
groups.\cite{nguyen11a,matthiesen12a,proux15a,arXiv_alkhuzheyri16a} It
is counter-intuitive as one would would expect a scattered photon to
retain also the second-order coherence of its laser, that is, to
remain uncorrelated with other scattered photons.  Note, however, that
it would be incorrect to state that the scattered photons alone are
antibunched. The effect is more subtle, and requires the coherent
absorption and re-emission process that originates from the
compensating Lorentzian lines, even though the total intensity cancels
out. Namely, filtering the emission to keep only the~$\delta$ peak
spoils the antibunching.\cite{gonzaleztudela13a} This is where lies
the counter-intuitive indeed characteristic of this emission: it
requires the contribution of the incoherent fluorescence even though
its intensity can be made vanishingly small! Our last assertion can be
easily checked experimentally: the Rayleigh peaks from the same laser
that excites two two-level systems with different lifetimes will
produce photons with different antibunching. For an ideal laser, both
peaks are the same~$\delta$ function. Yet, their antibunching is
different, proof that some aspect of the incoherent fluorescence is
involved, despite not explicitly in the power spectrum.

As far as strong-coupling is concerned, the situation of interest is
that of energy splitting between the dressed states. This occurs when:
\begin{equation}
  \label{eq:oldcrit}
    \Omega >\gamma_\sigma/8\,,
\end{equation}
and is shown in Figs.~\ref{Fig:fig2}(a) and~(c). Note that,
strikingly, the splitted states have a negative weight, i.e.,
correspond to absorbing lines. While their position is consistent with
the conventional dressed-state picture with
energies~$\omega_{\pm}=\pm \sqrt{64\Omega^2 -\gamma_\sigma^2}/4$, the
dynamics involved is opposite to that of new states radiating at their
corresponding energy. Instead, they enter the scene by absorbing
energy. They start to emit light instead of taking it away for the
more stringent condition (cf.~Eq.~(\ref{eq:oldcrit})):
\begin{equation}
  \label{eq:newcrit}
    \Omega >\gamma_\sigma/\sqrt{8}\,,
\end{equation}
Figure~\ref{Fig:fig2}(d) shows the emission spectrum for
$\Omega=\gamma_\sigma$ with, in this case, an active contribution from
all the dressed states that result in the characteristic emission
spectrum as a triplet that is clearly understood from the dressed
state structure. In the limit of very large intensity, $D_\pm \approx
3\gamma_\sigma/4 \pm 2 i \Omega$, and the emission spectrum is the
well-known Mollow triplet:\cite{mollow69a}
\begin{widetext}
  \begin{multline}
    \label{eq:mollow}
    S(\omega) =\frac{1}{2\pi}\frac{\gamma_\sigma/2}{(\gamma_\sigma/2)^2+\omega^2} + \frac{1}{4\pi}\frac{3\gamma_\sigma/4}{(3\gamma_\sigma/4)^2+(\omega+\chi/4)^2} \left[\frac{8\Omega^2-\gamma_\sigma^2}{8\Omega^2+\gamma_\sigma^2} +\frac{\gamma_\sigma(40\Omega^2+\gamma_\sigma^2)}{\chi (8\Omega^2+\gamma_\sigma^2)}\left(\omega+\frac{\chi}{4}\right)\right] +{}\\
    {}+\frac{1}{4\pi}\frac{3\gamma_\sigma/4}{(3\gamma_\sigma/4)^2+(\omega-\chi/4)^2}
    \left[\frac{8\Omega^2-\gamma_\sigma^2}{8\Omega^2+\gamma_\sigma^2}
      -\frac{\gamma_\sigma(40\Omega^2-\gamma_\sigma^2)}{\chi
        (8\Omega^2+\gamma_\sigma^2)}\left(\omega-\frac{\chi}{4}\right)\right]
    \,,
  \end{multline}
\end{widetext}
where $\chi = \sqrt{64\Omega^2-\gamma_\sigma^2}$. The satellites
contribute negatively to the total spectrum when they are close to the
center, with effect of trimming the fat tails of the central
Lorentzian. As a result, the system emits with a sharper distribution
of frequencies, typically a Student~$t$ distribution (that results
from the difference of two Lorentzians, as shown in
Fig.~\ref{Fig:fig2}(b)). This allows to collect more easily all the
emitted photon in a narrower window of detection and as a result to
achieve better values of antibunching. On the opposite, when the
Mollow triplet is largely split, it consists of essentially three
non-overlapping Lorentzian lines (of linewidths~$3\gamma_\sigma/4$
and~$\gamma_\sigma$ for the central and satellite peaks,
respectively), its statistical properties recover those of an
incoherently pumped~2LS.  This understanding of the composition of the
Mollow triplet, even before it is fully-formed, is important not only
on fundamental grounds, but also since it can be used to engineer
better single-photon sources, as we will see in
Section~\ref{sec:juefeb25200943CET2016}.

\subsection{Excitation by a single-photon source}
\label{sec:viemay13192349CEST2016}
\begin{figure*}
  \includegraphics[width=1\linewidth]{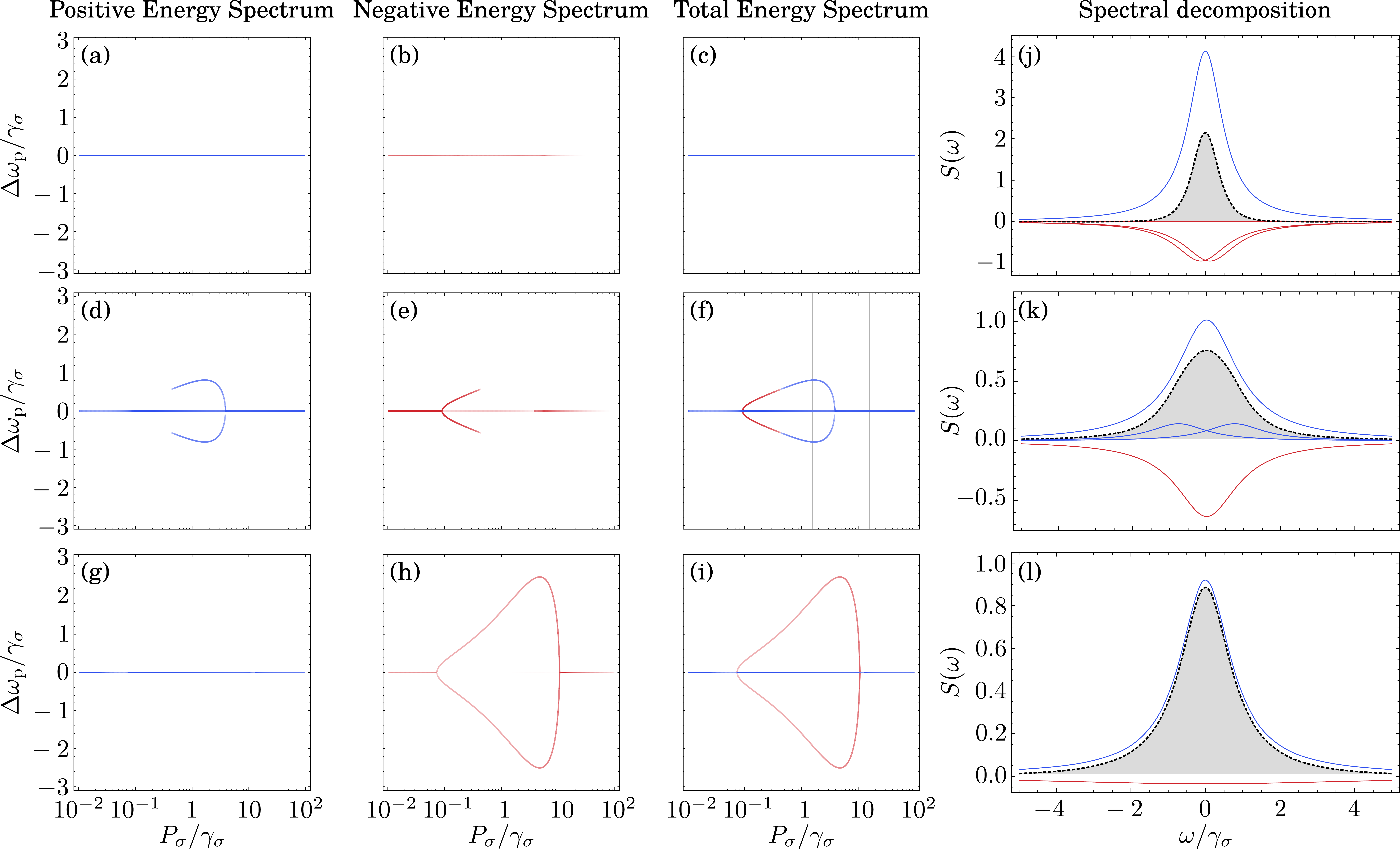}
  \caption{(Color online). Excitation by a single--photon
    source. (a-i)~Energy spectra of the target 2LS. The energies that
    contribute to the total emission spectrum with positive
    lorentzians are shown in blue, whereas those that contribute with
    negative lorentzians are shown in red. (j-l)~Emission spectra of
    the target 2LS (dashed, black lines) for the values
    $P_\sigma/\gamma_\sigma$ marked by the vertical dashed lines in
    Panel~(f). The total emission spectrum is made of the sum of
    positive (blue lines) and negative (red lines) lorentzians. To
    compute the figures we set $\gamma_\sigma$ as the unit, and set
    $\gamma_\xi/\gamma_\sigma=0.1$ in Panels~(a-c),
    $\gamma_\xi/\gamma_\sigma=1$ in Panels~(d-f), and
    $\gamma_\xi/\gamma_\sigma=10$ in Panels~(g-i).  }
  \label{fig:fig3}
\end{figure*}

In this and the following four Sections, we upgrade the source of
excitation from a classical to a quantum field. The simplest quantum
field is that provided by a single-photon source, which is modeled by
an incoherently driven 2LS, as sketched in Fig.~\ref{fig:fig1}(a) with
a thermal field (incoherent pumping) providing the classical
excitation, and is described by Eq.~(\ref{Eq.MasterEquation}) adding
the lindblad term $(P_\sigma/2)\mathcal{L}_{\ud{\sigma}}\rho$. The
resulting Liouvillian matrix $M$ has dimension $16\times 16$ but only
four of the sixteen eigenvalues contribute to the steady-state
emission spectrum. These eigenvalues are solutions to cubic equations
that are too bulky to be written here. In Fig.~\ref{fig:fig3}(a-i), we
show the weighted energy spectrum of the target 2LS as a function of
the intensity of the incoherent driving. For clarity, we split the
positive (1st column) and negative (2nd column) contributions that are
shown together in the the third column, with the same color code
(blue, positive; red, negative). The upper row,
Figs.~\ref{fig:fig3}(a-c), shows the case where
$\gamma_\xi/\gamma_\sigma=0.1$ and the driving is too weak to dress
the energies of the target 2LS: it behaves like its classical
counterpart under incoherent pumping.  Figure~\ref{fig:fig3}(d-f)
shows the case where $\gamma_\xi /\gamma_\sigma =1$, that results in a
splitting of the energy levels, with a splitting that opens with
absorbing lines before turning to emitting dressed states, as in the
case of a 2LS driven by a coherent classical field. Unlike the latter,
however, the splitting eventually quenches with increasing pumping
rate. This can be explained by the fact that the target 2LS is not
driven efficiently in the high excitation regime since the emission
spectrum of the source 2LS broadens (cf. Eq.~(\ref{eq:PLIncoherent2LS})),
reducing the intensity at the frequency of the target 2LS.  In
Fig.~\ref{fig:fig3}(f), we select three values for
$P_\sigma/\gamma_\sigma$ for which the total emission spectrum (dashed
black line) is decomposed into its dressed state emission (emitting in
blue and absorbing in red), as shown in Fig.~\ref{fig:fig3}(j-l).  The
magnitude of the lorentzians at the splitted energies is not large
enough to result in an observable splitting in the total emission
spectrum of the target 2LS.  Figure~\ref{fig:fig3}(g-i) shows the case
where $\gamma_\xi/\gamma_\sigma=10$, with a third scenario of an
energy splitting that occurs \emph{only} with absorbing dressed
states, but with a magnitude of the splitting that is larger than in
the previous case. This case also fails to produce an observable
splitting in the total luminescence spectrum. Even though the power
spectrum remains a single-peak throughout, one can still dress a 2LS
with a SPS.  The energy splitting of the target 2LS occurs when the
following condition is satisfied:
\begin{widetext}
  \begin{equation}
\label{Eq.ConditionSC}    
  \gamma_\xi/\gamma_\sigma \geq\left [ \frac{P_\sigma^{\ast 4}-68
      P_\sigma^{\ast 3}+726 P_\sigma^{\ast 2}+1-
      8\sqrt{-2P_\sigma^\ast \left(P_\sigma^{\ast 2}-
          16P_\sigma^\ast+1\right)^3}}
    {P_\sigma^{\ast 2}-14P_\sigma^\ast+1}\right]^{1/2}\, ,
\end{equation}
\end{widetext}
where we have introduced the unitless
parameter~$P_\sigma^\ast = P_\sigma/\gamma_\sigma$, and we note that
the condition is valid only when the argument on the rhs of the
inequality is a real value. The lower bound for~$P_\sigma^*$ is the
real-valued zero that cancels the rhs of Eq.~(\ref{Eq.ConditionSC}),
which is the solution of a polynomial of order~8 for which we give a
numerical approximation. When the rhs of Eq.~(\ref{Eq.ConditionSC}) is
real, it ranges from zero to infinty, which translates to a condition
on the population of the source $0.93\gtrsim n_\sigma\gtrsim 0.88$.
Equation~(\ref{Eq.ConditionSC}) thus provides the criterion for
strong-coupling of a 2LS with the light emitted in the CW regime by a
SPS. While this shows this is possible, with the source 2LS close to
saturation, it also shows the conditions as required by the formula,
are not particularly enlightening.
\begin{figure}
  \includegraphics[width=1\linewidth]{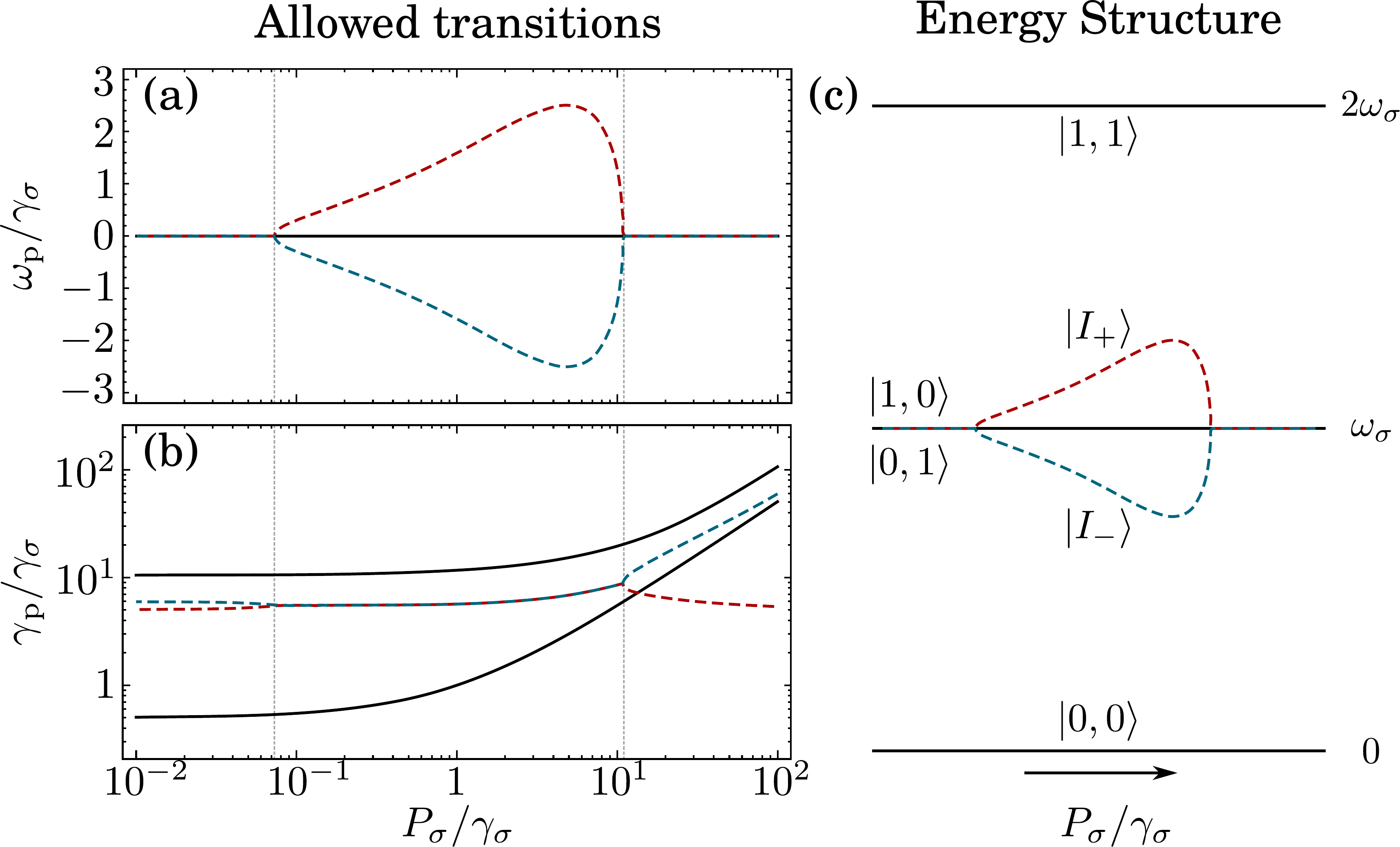}
  \caption{(Color online). Energy structure of a 2LS driven by a
    single photon source.~(a) Energy change in due to the allowed
    transitions in the 2LS.~(b) Linewidth of the emission of the 2LS
    due to the allwed transitions. The transitions that involved the
    dressed states are shown by the dashed red and dashed blue
    lines. Panel~(a) shows a splitting in the transition energies,
    while Panel~(b) shows that those transitions have the same
    linewidth.~(c) The energy structure of the 2LS driven by a single
    photon source is obtained from the allowed transitions in
    Panel~(a).  }
  \label{fig:fig4}
\end{figure}
Since the complex energy spectrum remains (relatively) simple, one can
go in this case one stage deeper and reconstruct from the resonances
the structure of the dressed states.

The archetype of the energy spectrum of a 2LS driven by a
single-photon source, and the corresponding linewidths are shown in
Fig.~\ref{fig:fig4}(a) and in Fig.~\ref{fig:fig4}(b), respectively.
The black solid lines in the figures correspond to transitions between
the bare modes, i.e., from $\ket{1}_\sigma\ket{1}_\xi=\ket{1,1}$ to
either $\ket{1,0}$ or $\ket{0,1}$, and from either $\ket{1,0}$ or
$\ket{0,1}$ to $\ket{0,0}$. The blue and red dashed lines correspond
to transitions involving dressed states that we note as $\ket{I_{\pm}}
= \alpha \ket{0,1} +\sqrt{1-\alpha^2} e^{\pm i \phi}\ket{1,0}$, where
$0\leq \alpha\leq 1$. However, the appearance of the dressed states do
not imply the absence of the bare state. On the contrary they coexist
giving rise to transitions from $\ket{1,1}$ to either $\ket{I_+}$ or
$\ket{I_-}$, but only from $\ket{0,1}$ to $\ket{0,0}$: when the energy
levels are splitted, the transition from $\ket{1,0}$ to $\ket{0,0}$ is
suppressed as all the emission of the source 2LS is efficiently
absorbed by the target 2LS, and instead takes place the coherent
transfer from $\ket{1,0}$ to $\ket{0,1}$. Therefore, the schematic
representation of the energy levels of a 2LS driven by a single-photon
source is as shown in Fig.~\ref{fig:fig4}(c), where we show the
coexistance of the bare and dressed states in the single-photon
manifold.

\subsection{Excitation by a Mollow triplet}
\label{sec:viemay13192407CEST2016}

\begin{figure*}
  \includegraphics[width=1\linewidth]{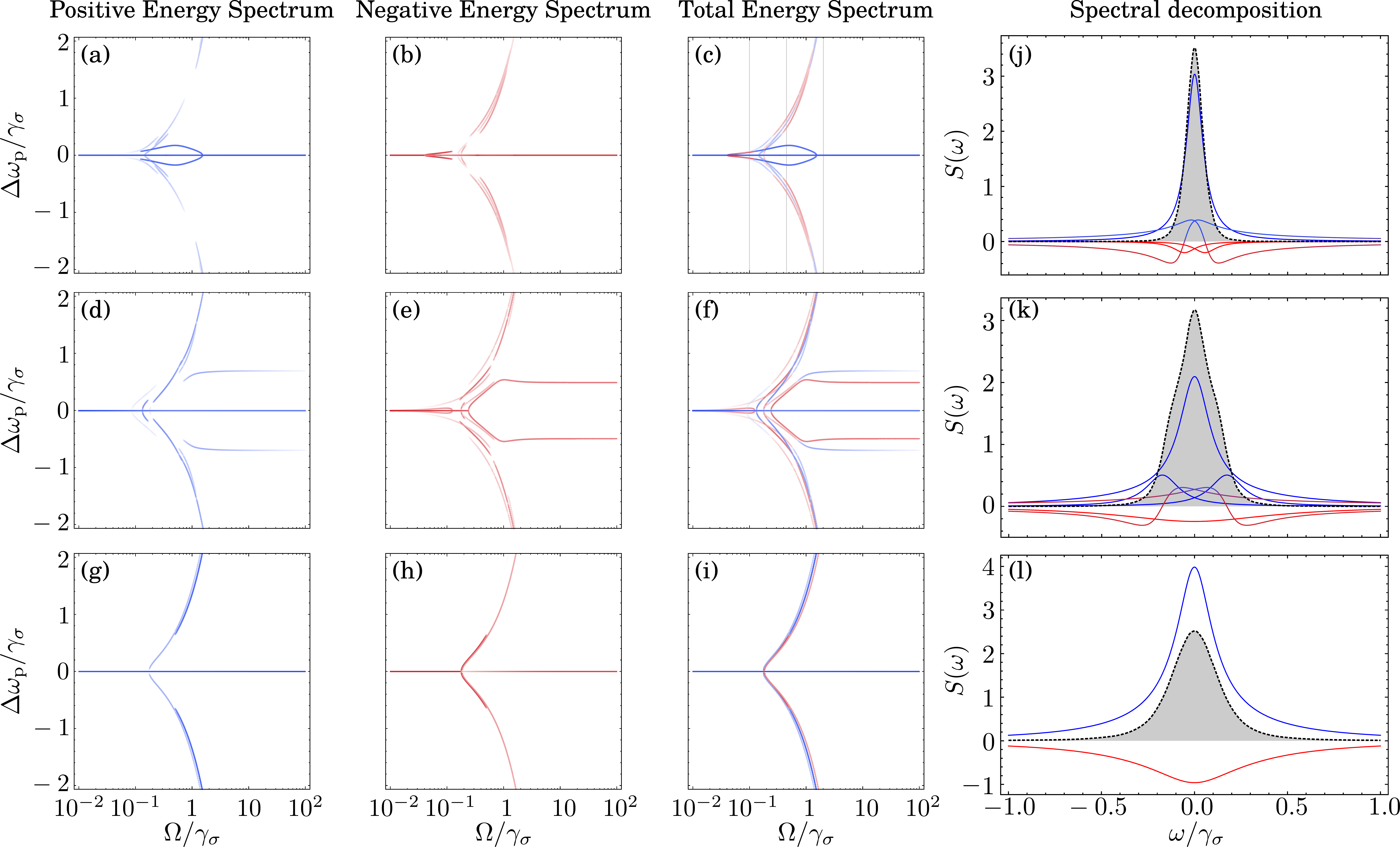}
  \caption{(Color online). Excitation by a Mollow
    triplet. (a-i)~Energy spectra of the target 2LS. The energies that
    contribute to the total emission spectrum with positive
    lorentzians are shown in blue, whereas those that contribute with
    negative lorentzians are shown in red. (j-l)~Emission spectra of
    the target 2LS (dashed, black lines) for the values
    $\Omega/\gamma_\sigma$ marked by the vertical dashed lines in
    Panel~(c). The total emission spectrum is made of the sum of
    positive (blue lines) and negative (red lines) lorentzians. To
    compute the figures we set $\gamma_\sigma$ as the unit, and set
    $\gamma_\xi/\gamma_\sigma=0.1$ in Panels~(a-c),
    $\gamma_\xi/\gamma_\sigma=1$ in Panels~(d-f), and
    $\gamma_\xi/\gamma_\sigma=10$ in Panels~(g-i). }
  \label{fig:fig5}
\end{figure*}

In this Section, we keep the same source 2LS but we excite it
coherently. This systematic study of all the possibilities of exciting
a 2LS by quantum light thus brings us to this curious configuration of
exciting a 2LS by the Mollow triplet.  The system is described by
Eq.~(\ref{Eq.MasterEquation}), using two input channels for the source
with $\epsilon_1=\epsilon_2=1/2$, setting
$H_\sigma = -i \sqrt{\epsilon_1\gamma_\sigma}\mathcal{E}
(\ud{\sigma}-\sigma)$, and for simplicity we note
$\Omega \equiv \sqrt{\epsilon_1\gamma_\sigma}\mathcal{E}$ as the
intensity of the coherent light that drives the source 2LS. As in the
previous section, the Hilbert space has dimension $16\times 16$, but
in this case all the eigenvalues are involved in the energy spectrum
of the target 2LS, as shown in Fig.~\ref{fig:fig5} for several values
of the ratio $\gamma_\xi/\gamma_\sigma$.  The most obvious observation
is the considerably higher complexity of the structure of the driven
system.  In this case, we will focus only on the resonances without
making any attempt at reconstructing the underlying dressed states, as
we did for the simplest configuration of exciting with a SPS.
Panels~(a-c) show the case for $\gamma_\xi/\gamma_\sigma=0.1$.  The
splitting also starts with absorbing states and occurs for a driving
intensity $\Omega/\gamma_\sigma \approx 0.02$, that is is roughly one
order of magnitude smaller than that required with a classical laser
($\Omega/\gamma_\sigma =1/8$). Other states do split at that usual
threshold, showing that the target 2LS incorporates the energy
structure of the source to its own.  In fact, at large driving
intensities, most of the energies converge to those of the source 2LS
to provide a carbon copy of the conventional Mollow triplet,
cf.~Fig.~\ref{Fig:fig2}.  The most intense contributions to the total
emission spectrum remain those provided by eigenvalues for which
$|\Delta\omega_{\mathrm{P}}/\gamma_\sigma|<1$.

The total emission spectrum (dashed black lines) and its decomposition
through dressed states emission (blue and red solid lines) are shown
in Fig.~\ref{fig:fig5}(j-l) for the values of $\Omega/\gamma_\sigma$
marked by the vertical dashed lines in panel~(c).  Panel~(j) displays
another case of emergence of the dressed states as absorbing lines,
beside, in this case, also strongly interfering with important
dispersive components (dotted lines). As a result, the total spectrum
broadens as compared to the simple target-2LS emission (blue line).
Panel~(k) shows how the dressed states now fully formed contribute to
the emission with enough energy splitting and weight of the
corresponding transition to produce a noticeable feature in the total
emission spectrum (dotted black). However, negative contributions
balance these peaks and the resulting emission spectrum only exhibits
two ``bumps'' on its flanks. The target is resilient to developing a
Mollow triplet, although its dressing is now unambiguous as observable
directly in the luminescence.  Panel~(l) shows the quenching of the
Mollow triplet at higher intensity of the source: the lateral peaks
vanish and the remaining splitted energies lie close to the bare
state, resulting in an emission spectrum with a single,
heavily-tailed, line.

Panels~(d-f) and~(g-i) of Fig.~\ref{fig:fig5} show the energy spectrum
of the target 2LS for $\gamma_\xi/\gamma_\sigma=1$ and
$\gamma_\xi/\gamma_\sigma=10$, respectively. Although most of the
features of the energy spectrum of the target 2LS remain when we
increase its decay rate, a notable feature is observed when
$\gamma_\xi/\gamma_\sigma=1$: there appears in this case dressed
states that are splitted in energy at all driving intensities, as seen
in Fig.~\ref{fig:fig5}(e) for $\Omega/\gamma_\sigma < 10^{-1}$ through
the asymptotic lines instead of the usual bifurcations. While the
weights also vanish with decreasing driving intensity, they are never
exactly zero and the structure of the dressed states is peculiar as
they manifest including at vanishing driving field intensity. This is
another manifestation of how equal decay rates in coupled
quantum-optical system lead to optimum strong-coupling
conditions.\cite{laussy12d} In this case, this opens a channel of
excitation where the target 2LS can benefit from the strong-coupling
of the source 2LS (that is in strong-coupling from the strong
classical driving field) regardless of its driving intensity. The
information of strong-coupling is therefore ``encoded'' in the photons
emitted by the source and ``restored'' in the target. If the target
and sources are different objects (due to mismached decay rates), this
information is lost.  

% As we keep on increasing the decay rate of the
% target 2LS its energy spectrum becomes more similar to that of the
% source 2LS (not shown).

\subsection{Excitation by an incoherently driven cavity}
\label{sec:viemay13191706CEST2016}

\begin{figure*}
  \includegraphics[width=1\linewidth]{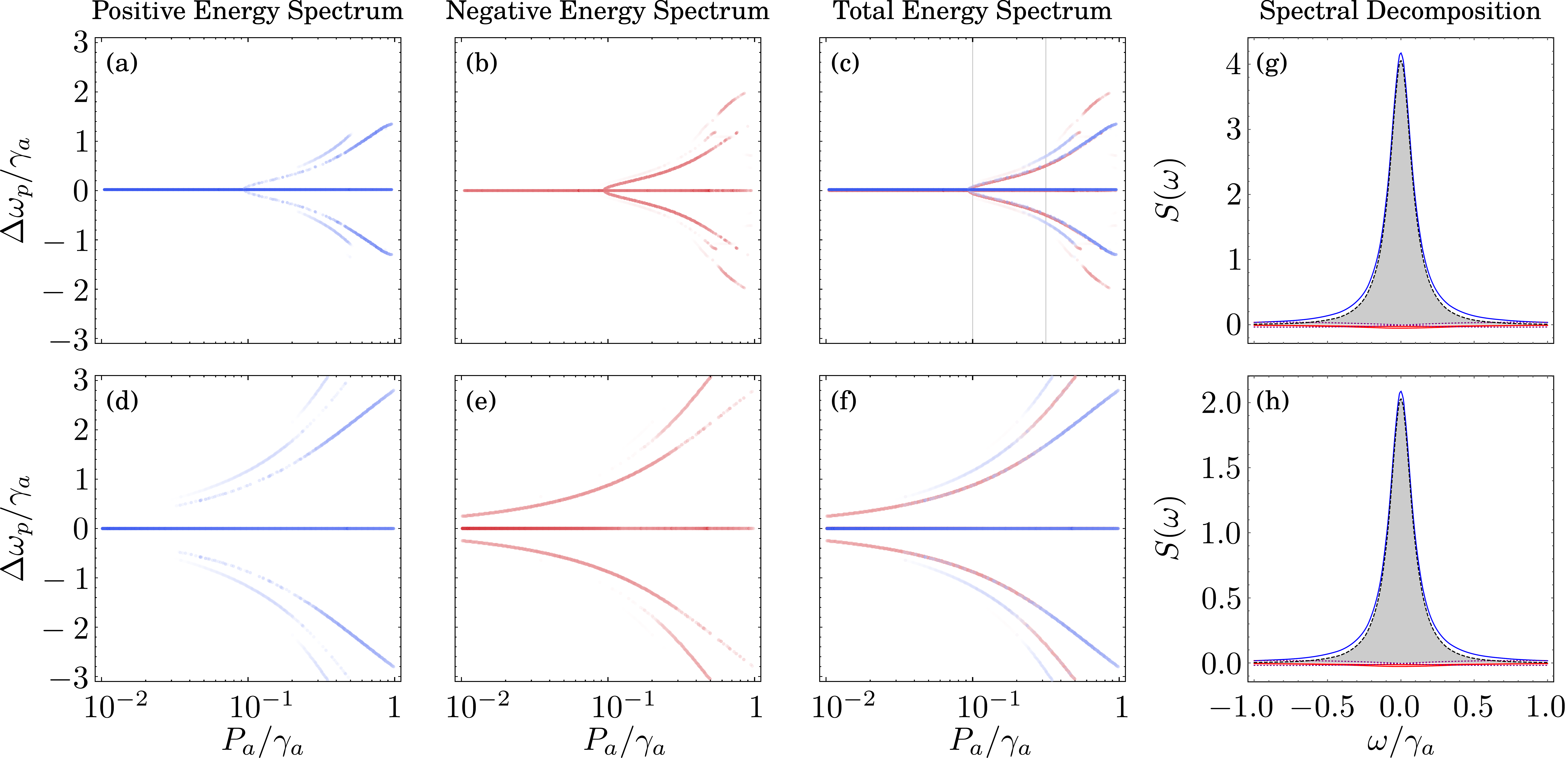}
  \caption{(Color online). Excitation by an incoherently driven
    cavity. (a-f)~Energy spectra of the target 2LS. The energies that
    contribute to the total emission spectrum with positive
    lorentzians are shown in blue, whereas those that contribute with
    negative lorentzians are shown in red. (g-h)~Emission spectra of
    the target 2LS (dashed, black lines) for the values
    $P_a/\gamma_\sigma$ marked by the vertical dashed lines in
    Panel~(c). The total emission spectrum is made of the sum of
    positive (blue lines) and negative (red lines) lorentzians. To
    compute the figures we set $\gamma_a$ as the unit, and set
    $\gamma_\xi/\gamma_a=0.1$ in Panels~(a-c), and
    $\gamma_\xi/\gamma_a=1$ in Panels~(d-f).}
  \label{fig:fig6}
\end{figure*}

In this and the following Section, we change the source of excitation
for a driven cavity. Here, we discuss the case of an incoherently
driven cavity. The system is described by
Eq.~(\ref{Eq.MasterEquation}) when replacing the parameters of the
source (operators and associated variables), marked with the subindex
$\sigma$, by those corresponding to a cavity, which we note with the
subindex $a$ (the operator~$a$ is a boson annihilation operator).  The
incoherent driving of the cavity is described by the lindblad term
$(P_a/2)\mathcal{L}_{\ud{a}} \rho$ in the master equation. In contrast
to the previous three Sections, the Hilbert space of the system is now
infinite. To compute the energy spectrum, we need to truncate in the
number of excitations. This is a difficult constrain for an
incoherently driven cavity since its population
$n_a = P_a /(\gamma_a -P_a)$ has thermal fluctuations.% In order for
% the propability to have $k$~photons in the cavity to be less than
% $10^{-\epsilon}$, $k$ has to satisfy
% $k > -\frac{\epsilon + \log (1-P_a/\gamma_a)}{\log(P_a/\gamma_a)}$ which 
% %
% For example, if we want to have a cavity with $9$~photons (or
% equivalentlty with $P_a/\gamma_a=0.9$) such that the probability to
% have $k$~photons is less than $10^{-5}$, we need $k>88$ so the
% liuvillian matrix would be of dimension $180\times 180$, and therefore
% hard to diagonalize exactly.
For this reason the energy spectra showed in Fig.~\ref{fig:fig6} are
computed up to $P_a/\gamma_a \approx 0.968$, with corresponding
populations $n_a=30$ for which we have checked convergence of the
results.  Panels~(a-c) display the case with $\gamma_\xi/\gamma_a=0.1$
and show that the energy spectrum is qualitatively similar to a 2LS
driven by a classical laser (cf.~Fig.~\ref{Fig:fig2}(a)). The most
notable features of this result are the following: first, weighting
the transitions in a truncated Hilbert space is mandatory to reach
convergence, as $D_p$ alone do not yield a stable structure otherwise
as not only the number of eigenvalues grows with the size of
truncation, but their distribution also fails to settle to a
consistent pattern.  Second, the weighted complex energy spectrum,
shown in Fig.~\ref{fig:fig6}, that is well-defined and for which we
have checked convergence, reduces to a simple structure in the
spectral shape but still exhibits some unexpected features, namely, up
to three satellite energies visible on each flank of the central
Lorentzian, as compared to the single line of the conventional Mollow
triplet (cf.~Fig.~\ref{Fig:fig2}(a)). The system fluctuates so much
that it cannot accommodate all the transitions with three energies
only.  Third, while the weighted structure is indeed similar to the
classical excitation, the intensity of the satellite peaks is much
smaller in comparison, so that the total emission spectrum is barely
affected by them and is given essentially by a single lorentzian at
the energy of the target 2LS, as shown in Fig.~\ref{fig:fig6}(g-h) for
the values $P_a/\gamma_a$ marked by the vertical dashed lines in
Fig.~\ref{fig:fig6}(c). The excitation of a 2LS by thermal light is
therefore largely a case of fundamental interest that, for most
purposes, can be approximated to an incoherently excited 2LS, despite
this configuration yields a dressing of the target. Fourth, as in the
previous case, setting the decay rate of the target 2LS equal to the
decay rate of the cavity leads to energy-splitting at all values of
pumping, while a mismatch leads to a bifurcation instead, as shown in
Fig.~\ref{fig:fig6}(d-f). In summary for this configuration, there is
little physics in the main observable, but much to be learned on the
mechanisms of quantum excitation from a (quantized) thermal source.

\subsection{Excitation by a coherently driven cavity}
\label{sec:TueAug30175459CEST2016}
\begin{figure*}
    \includegraphics[width=1\linewidth]{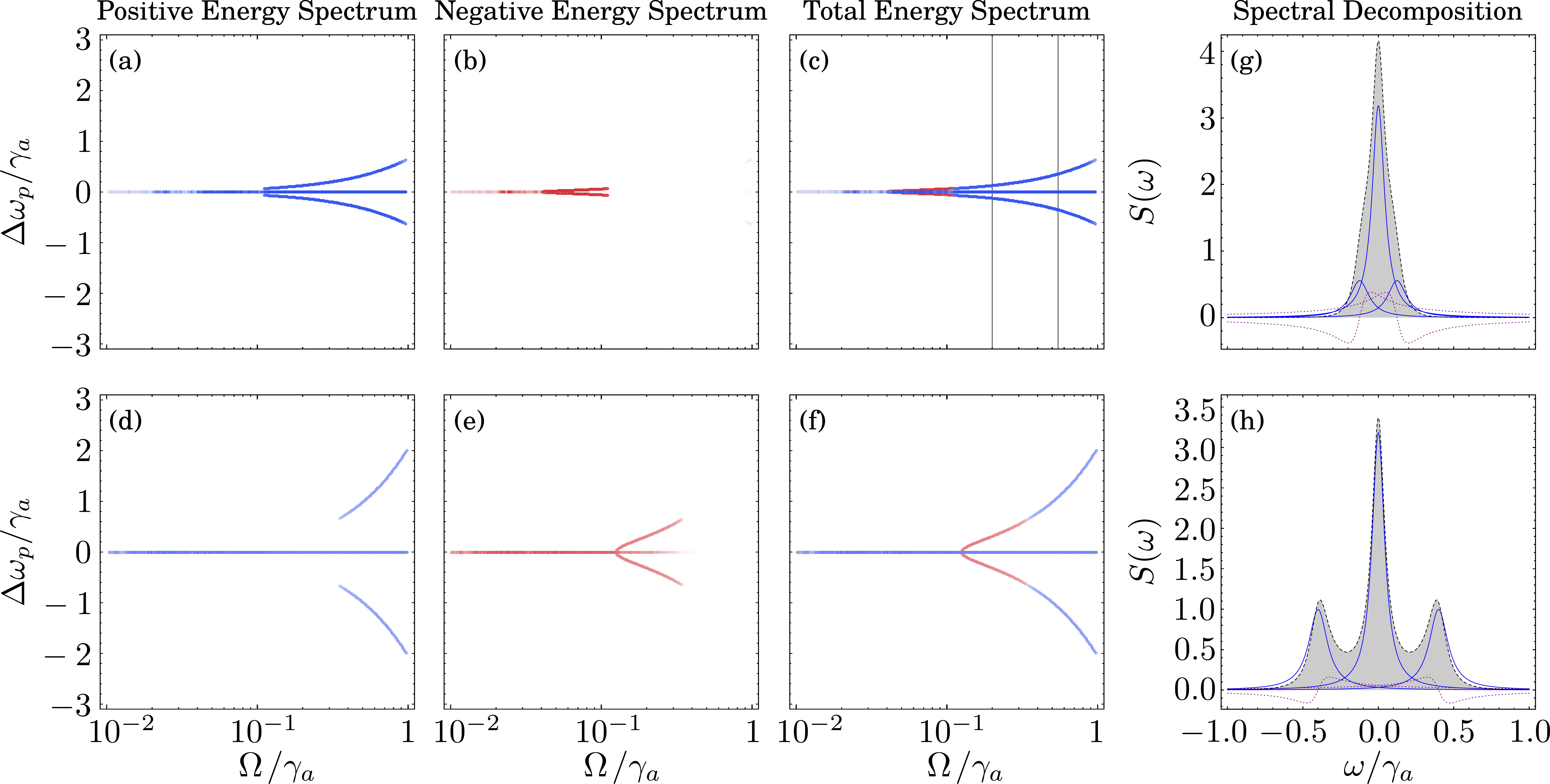}
    \caption{(Color online). Excitation by a coherently driven
      cavity. (a-f)~Energy spectra of the target 2LS. The energies
      that contribute to the total emission spectrum with positive
      lorentzians are shown in blue, whereas those that contribute
      with negative lorentzians are shown in red. (g-h)~Emission
      spectra of the target 2LS (dashed, black lines) for the values
      $\Omega/\gamma_\sigma$ marked by the vertical dashed lines in
      Panel~(c). The total emission spectrum is made of the sum of
      positive (blue lines) and negative (red lines) lorentzians. To
      compute the figures we set $\gamma_a$ as the unit, and set
      $\gamma_\xi/\gamma_a=0.1$ in Panels~(a-c), and
      $\gamma_\xi/\gamma_a=1$ in Panels~(d-f).}
  \label{fig:fig7}
\end{figure*}

In this Section, we describe the energy splitting of a 2LS driven by
the emission of a coherently driven cavity.  The system is described
by Eq.~(\ref{Eq.MasterEquation}) with the changes of the previous
Section, but now we use two input channels for the source with
$\epsilon_1=\epsilon_2=1/2$, we replace the incoherent pumping of the
source by $H_a= -i \sqrt{\epsilon_1\gamma_a}\mathcal{E} (\ud{a}-a)$,
and set the effective driving intensity of the source as
$\Omega=\sqrt{\epsilon_1\gamma_a}\mathcal{E}$. 
This creates a coherent state~$\ket{2\Omega/\gamma_a}$ that is
dynamically coupled to the 2LS. The dimension of the Hilbert space is
also infinite and must be truncated. In this case, however, Poissonian
fluctuation of the cavity make the truncation manageable up to large
populations.
% $n_a = (2\Omega/\gamma_a)^2$, the truncation required to account for
% $99.9\%$ of the photons we need $k > n_a +3\sqrt{n_a}$. Thus, for
% $n_a =4$ (or equivalently $\Omega/\gamma_a=1$) we need a truncation of
% 20 photons, and the liuviilian matrix can be easily diagonalized.

The energy spectrum of the target 2LS are shown in
Fig.~\ref{fig:fig7}(a-c) for $\gamma_\xi/\gamma_a=0.1$, and in
Fig.~\ref{fig:fig7}(d-f) for $\gamma_\xi/\gamma_a=1$. The notable
features here are direct counterparts of the incoherent excitation
case: First, the weighted energy spectrum now fully reduces to the
level of complexity of the conventional Mollow triplet, with exactly
three resonances. Here it must be borne in mind that the underlying
structure is that of a fully quantized 2LS-cavity system, with
countably infinite complex eigenvalues (that fail to provide a
converged structure if unweighted). In the Hamiltonian regime, the
corresponding structure is extremely complicated~\cite{delvalle11a}
and this is the cascaded feature (of no feedback from the target to
its source) that brings such a simplification with an actual triplet
structure down to the most fundamental level. This is not exactly the
conventional Mollow case, however, since there is one additional
parameter, the decay rate of the cavity~$\gamma_a$, that cannot be
zero as otherwise no photons are emitted by the source. This causes
some differences between these two configurations: the $c$-field on
the one-hand and the coherent-state cavity on the other.
Figure~\ref{Fig:fig2} shows two cases for the later with
different~$\gamma_\xi/\gamma_a$.  Interestingly, in this case, the
configuration~$\gamma_a=\gamma_\xi$ does not lead to an asymptotic
loss of the splitting but to a bifurcation. This splitting from a
single-line always occur before that of the conventional Mollow
triplet. The energy spectra remain qualitatively very similar to that
of the $c$-field driving, cf.~Fig.~\ref{Fig:fig2}. The agreement is
recovered at large drivings where a mean-field approximation, that
replaces the cavity operator $a$ by $\sqrt{n_a}$ in
Eq.~(\ref{Eq.MasterEquation}), converges to the numerical
solution. This approximation leads to the master equation of a 2LS
driven by a laser with
intensity~$\Omega^\ast=\sqrt{(1-\epsilon_1)\gamma_\xi \gamma_a
  n_a}=\sqrt{\gamma_\xi \gamma_a n_a/2}$
so the agreement at large~$\Omega$ (where $n_a$ is very large too)
appears to be exact. The mean-field approximation does not hold
however for small values of $\Omega/\gamma_a$ and therefore fails to
predict the splitting threshold. In this case, strong correlations
between the 2LS and the few photons from the quantized driving field
(albeit in a coherent state) exist that lead to differences from the
$c$-field driving. We come back to their importance in the last part
of the paper, Section~\ref{sec:juefeb25200943CET2016}, where we turn
to applied considerations of this physics.

\subsection{Excitation by a one-atom laser}
\label{sec:viemay13191827CEST2016}

\begin{figure*}
  \includegraphics[width=1\linewidth]{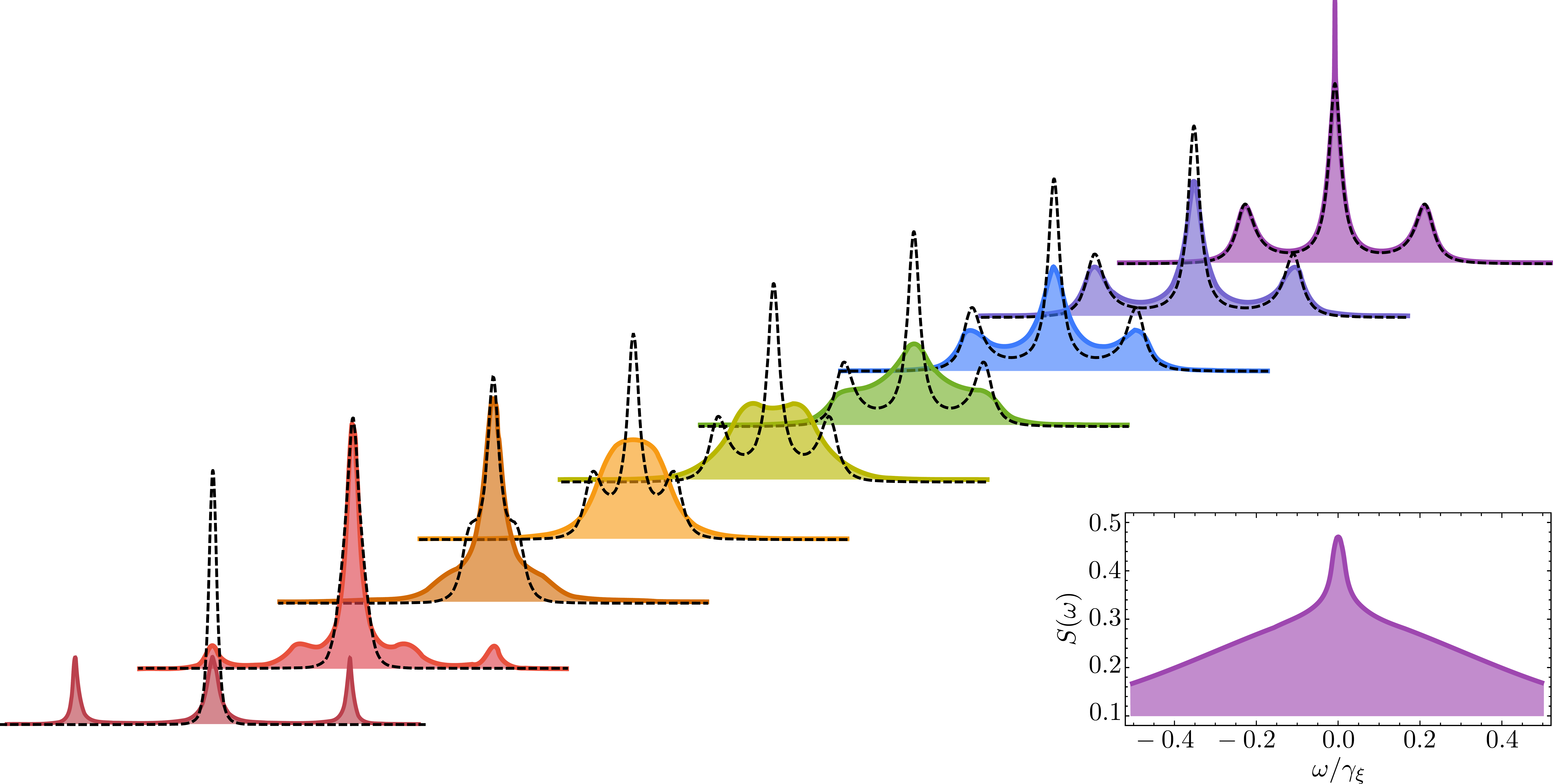}
  \caption{(Color online). Dressing of the target 2LS with a one atom
    laser. We compare the normalised emission spectra of the a 2LS
    driven by the emission of a one atom laser (solid color lines)
    with the normalised emission spectra of a 2LS driven coherently by
    a classical laser (dashed black lines). The incoherent driving of
    the one atom laser increases towards the upper right corner of the
    figure, and makes the emission spectra of the target 2LS to change
    from a filter of  the emission of the cavity, as is clear from the
    lateral peaks due to the strong coupling between the cavity and
    the atom inside, to the splitting of the emission line until it
    reaches a triplet shape equivalent to that of a coherently driven
    2LS.  To compute the figures we set the atom, the cavity and the
    target 2LS in resonance, and the rest of the parameters were as
    follows: $\gamma_\xi$ was set as the unit,
    $\gamma_\sigma/\gamma_\xi=10^{-2}$, $\gamma_a/\gamma_\xi=1$,
    $g/\gamma_\xi=10$. The incoherent driving rate $P_\sigma$ is
    different for each spectrum, for the spectrum at the bottom we
    took the limit $P_\sigma/\gamma_\xi \rightarrow 0$.  For the next
    seven lines we used the $P_\sigma$ such that
    $\Omega/\gamma_\xi=\sqrt{\gamma_a \gamma_\xi n_a}/\gamma_\xi=0.5$,
    $1.0$, $1.5$, $2.0$, $2.5$, $3.0$, and $3.5$. The last spectrum,
    at the top right of the figure, was obtained for a cavity with
    $\gamma_a/\gamma_\xi =0.1$ and $P_\sigma/\gamma_\xi=20.115$, so
    that the population in the cavity was $100$ photons. The inset
    shows a zoom on this last spectrum very close to the resonance of
    the target 2LS, to show the Rayleigh peak due to the driving by a
    laser. }
  \label{fig:fig8}
\end{figure*}

In this Section, we consider an actual laser as the source for the
2LS.  Namely, we drive the 2LS with the coherent state generated by a
device that creates this coherence from its internal dynamics, without
inheriting it from another classical source (e.g., another laser, a
classical field, etc.) This is the culminating point of our
description of the Mollow physics in this text as there is no bad
approximation that results in unphysical and/or pathological results,
which can result from the~$\delta$ peak of the
laser.\cite{gonzaleztudela13a} There have been several efforts to go
beyond the Mollow paradigm where the driving light is a perfect sine
wave, for instance Zoller's earliest work (that was the topic of his
Ph.~D thesis)\cite{zoller78a,zoller78b,zoller78c}, where the sought
features of the driving field are enforced into the model. Here, in
line with the cascaded formalism philosophy, we leave to the source to
self-consistently develop and establish its lasing properties.  The
simplest such laser is the one-atom laser,\cite{mu92a} that is able to
turn an incoherent pumping of the 2LS into a coherent state of the
cavity.\cite{mckeever03a,nomura10a,arXiv_gericke16a} Its Hamiltonian
is that of Jaynes and Cummings:\cite{jaynes63a}
\begin{equation}
  H_s = \omega_\sigma \ud{\sigma}\sigma + \omega_a
  \ud{a}a + g(\ud{a}\sigma + \ud{\sigma}a)\, ,
\end{equation}
where, as before, $\sigma$ is the fermionic operator describing the
atom and $a$ is the bosonic operator describing the cavity, with
respective free energies $\omega_\sigma$ and $\omega_a$. Now, however,
they are coupled reversibly with strength~$g$.  The incoherent driving
of the atom ($P_\sigma$), the decay processes and the cascaded
coupling between the cavity and the target 2LS are included in the
master equation:
\begin{multline}
  \partial_t \rho = i[\rho, H_s + H_\xi] + \frac{P_\sigma}{2}
  \mathcal{L}_{\ud{\sigma}} \rho+\sum_{k=\lbrace
    \sigma,a,\xi\rbrace} \frac{\gamma_k}{2}\mathcal{L}_k \rho 
  + {}\\
  {} + \sqrt{\gamma_a \gamma_\xi} \left \lbrace 
    [a \rho, \ud{\xi}] + [ \xi,\rho \ud{a}] \right \rbrace\, .
\end{multline}

Here, $\gamma_\sigma$, $\gamma_a$ and $\gamma_\xi$ are the decay rates
of the atom inside the cavity, the cavity, and the target 2LS,
respectively; and since there is no external coherent field driving
the system we consider only one input channel for the source and for
the target. In Fig.~\ref{fig:fig8} we show the normalised emission
spectra of a 2LS cascaded by a one atom laser (solid lines), and we
compare it to the normalised emission spectra of a 2LS driven by a
classical laser (dashed lines). In the bottom three lines of
Fig.~\ref{fig:fig8}, the incoherent driving of the atom is so small
that the one atom laser is not in the lasing regime. Therefore, the
comparison of the spectra is made so that the population in the target
2LS and in the coherently driven 2LS are the same. In all the other
cases, when the one atom laser is in the lasing regime, the comparison
of the spectra is made so that the driving intensity is the same,
i.e., we compare the spectra of the target 2LS with that of a 2LS
driven coherently with intensity $\Omega^\ast = \sqrt{\gamma_\xi
  \gamma_a n_a}$, where $n_a$ is the population of the cavity.
Figure~\ref{fig:fig8} shows neatly the splitting of the energy levels
of the target 2LS as a function of the rate of incoherent driving of
the atom inside the cavity. At the lowest $P_\sigma$, the spectrum of
the target 2LS is the one at the left bottom part of the figure. There
we see clearly that the target 2LS is acting as filter of the emission
of the cavity: the central peak correspond to the resonance of the
target 2LS, while the other two peaks reveal the Rabi doublet due to
the strong coupling between the cavity and the atom inside. As the
incoherent driving of the atom increases, the peaks corresponding to
the Rabi doublet become less dominant, and the emission spectrum of
the target 2LS tends to merge into a broad single line. Increasing
even further the incoherent driving, the emission line of the target
2LS begins to split again, but in a different way than the coherently
driven 2LS: while the splitting of the target 2LS seems to be a
doublet as e.g. in the light green line, the spectrum of the
coherently driven 2LS has clearly a triplet shape. At large driving
rates the emission of the target 2LS converges to that of the
coherently driven 2LS.  In fact, in the uppermost spectrum the cavity
has 100 photons, its statistics is coherent (that is
$\mean{\ud{a}\ud{a}aa}/n_a^2=1$), and the two spectra are exactly the
same. 

\begin{figure*}[t]
    \includegraphics[width=\linewidth]{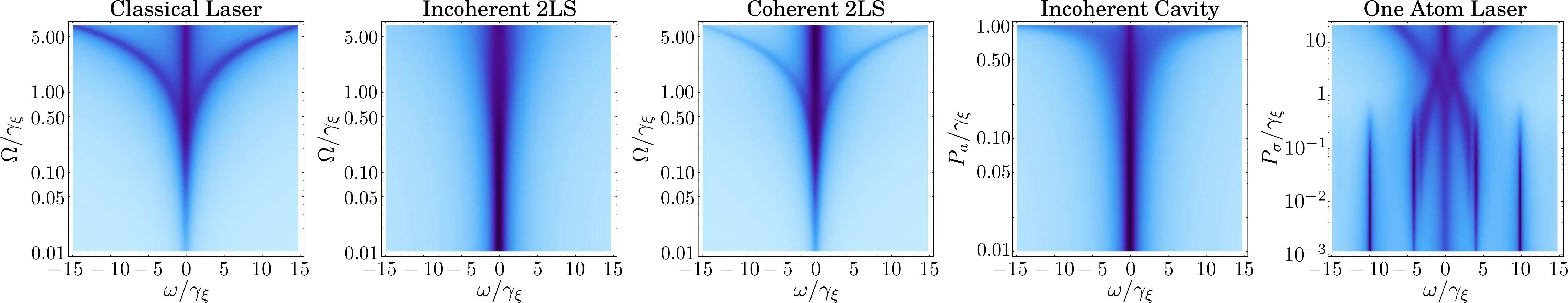}
    \caption{(Color online). Photoluminescence spectra of the
      two-level system driven by all the sources studied in the text:
      (a) A classical laser resulting in the conventional Mollow
      Triplet. (b) An incoherent 2LS providing a single line that
      merely broadens. (c) A coherent 2LS providing a Mollow splitting
      but of weak intensity and that vanishes with increasing driving
      intensity. (d) A Thermal cavity providing a single line whose
      broadening is reminiscent of the Mollow structure. (e) A
      one-atom laser showing the transition from the non-lasing regime
      of the driving source to the exact recovering of the
      conventional Mollow triplet. At low driving, the target echoes
      the structure from a Jaynes-cummings-like coupling with the
      source.}
  \label{fig:MollowDensity}
\end{figure*}

In Fig.~\ref{fig:MollowDensity} we show a density plot of the
photoluminiscence spectrum of the target 2LS driven by all the sources
of light that we have considered in this Section, and summarized in
its caption.

\section{Cascaded Single Photon Sources}
\label{sec:juefeb25200943CET2016}

Cascading is a powerful concept that allows to
achieve extremely high end values from a moderate initial input, with
such compelling examples as the domino effect~\cite{whitehead83a} that
can amplify energy by over two billions in a basement, to trophic
cascade~\cite{lindeman42a} that can lead to extinction of species.  It
acts in several key processes of various areas of science, e.g., with
chemical~\cite{bodenstein13a} and nuclear~\cite{anderson39a} chain
reactions.  In optics, through stimulated emission, it underpins basic
phenomena such as supperradiance~\cite{dicke54a} and
lasing.\cite{schawlow58a} With the advent of heterostructures, it
became possible to engineer more elaborate schemes to better control
the chain reaction. An highlight is the proposal by Kazarinov and
Suris of energy staircases in a superlattice leading to the successive
creations of an increasing number of photons by a single initial
electron,\cite{kazarinov71a} a scheme realized a quarter of a century
later under the name of a quantum cascade
laser.\cite{faist94a} Recently, another proposal was made with
cascades between condensates~\cite{liew13b} and extremely high
correlations in the form of superbunching were shown to occur as a
result of the cascading.\cite{liew16a}

\subsection{Driving the cascaded SPS}

In this Section, we propose the application of such a principle to
quantum light, and with the aim of increasing not the intensity of
light but a quality dear to quantum engineers, namely, the suppression
of multiple-photon emission, known as antibunching.\cite{paul82a}
This goal is highly pursued to power quantum information processing,
with boson sampling~\cite{aaronson11a}---not the most useful but the
most accessible demonstration of quantum parallelism out of the
classical reach---already in sight provided one could power linear
optical setups with slightly better SPS. For this reason, there is a
race to build always better sources, in particular in the
semiconductor community where such devices would furthermore have a
large economic and technological
potential.\cite{rau14a,schlehahn15a,ding16a} The Fourier transform
limit for single photon emission has already been
reached~\cite{kuhlmann15a} and there is now much efforts to combine
and enhance other features such as brigthness, efficiency and, of
course,
antibunching.\cite{unsleber16a,somaschi16a,loredo16a}
Our proposal takes a new direction and rather than bettering
engineering and implementation, we turn to a different mechanism to
increase the quality of SPS by magnitudes not accessible only with a
better technology. Namely, we propose to cascade the output of a chain
of single photon sources (SPS), with effect of a profound
restructuring of their spectral emission, bringing the initial
Lorentzian shape of a single SPS to grow into a Student~$t$ lineshape
of order~$2k$ after~$k$ iterations of the cascade.  We show that such
a spectral engineering that trims the fat Lorentzian tails is
accountable for increased antibunching. As the iterations converge
towards a Normal distribution, our scheme allows to engineer extremely
antibunched single photon sources that could power quantum logic with
the repetition rates necessary for their successful operations at a
large scale. In essence, our results remove the constrain of spectral
broadening associated to short lifetimes. Stated otherwise, it
achieves at the single-photon emission level what a laser does in the
Schallow Townes limit by narrowing the line with increasing signal.

Our proposal consists of an array of consecutive SPS, in which the
first SPS is excited externally while the rest of them are excited by
the fluorescence of the previous SPS. We will consider cascades of up
to three emitters but the scheme can be continued indefinitely. The
underlying principle is that exciting with quantum light allows to
access new regimes that neither classical excitations nor reversible
(Hamiltonian) coupling can provide.\cite{arXiv_lopezcarreno16a} This
is thanks to the added degrees of freedom of quantum light on the one
hand (such as reduced fluctuations) and the one-way transfer of energy
on the other hand that removes the effective decay implied by
strong-coupling oscillating the excitation back to its source. 

\begin{figure}[b]
  \includegraphics[width=1\linewidth]{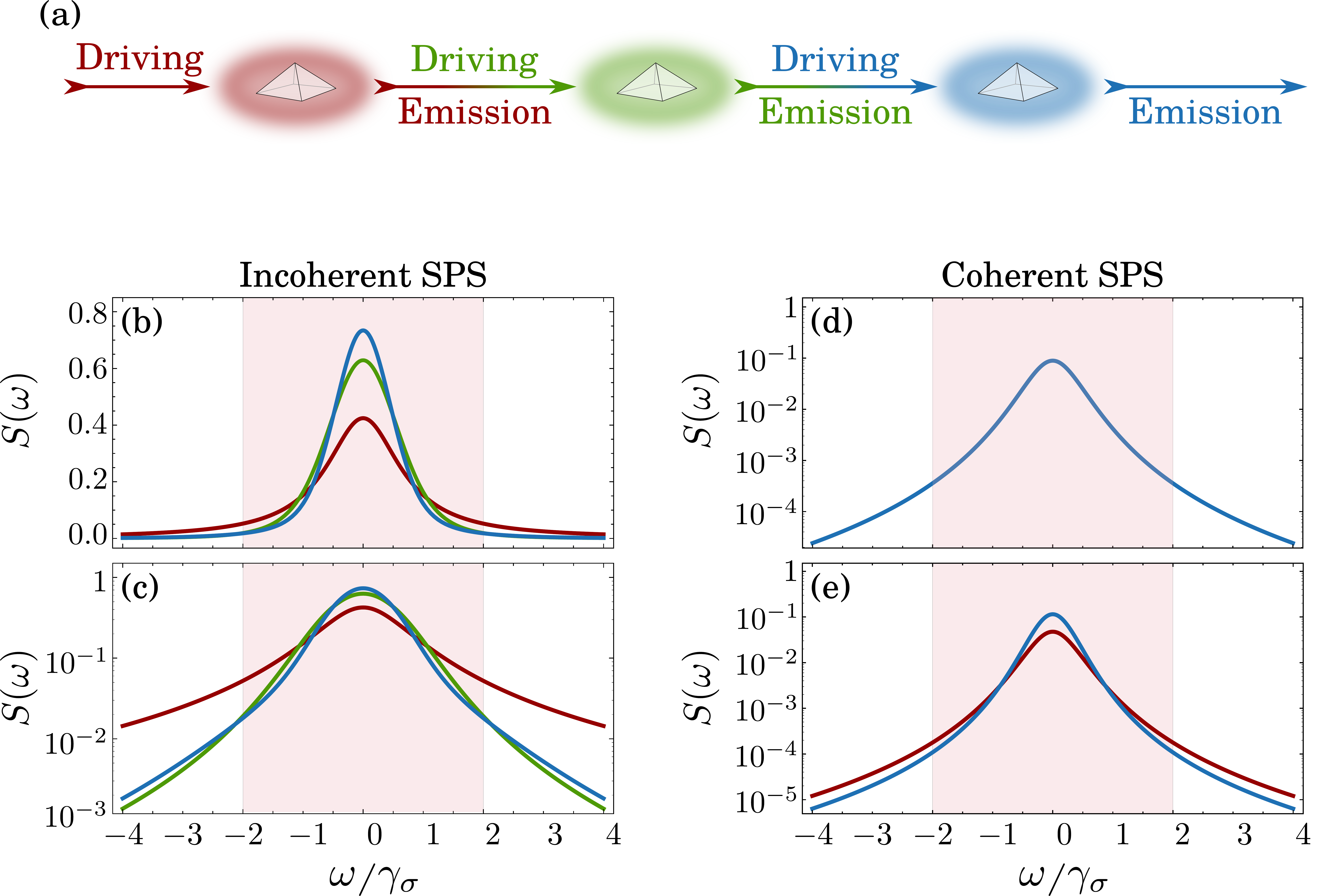}
  \caption{(Color online). Scheme of the cascaded single-photon
    sources. (a) A series of SPS is connected through a cascaded
    architecture. In each step of the cascade a SPS is driven by the
    emission of the previous SPS. (b--d) Emission spectra of the
    initial SPS (red), the first (green), and second (blue) cascade
    steps. In each step of the cascade the emission line becomes
    narrower and more peaked around the central frequency. Panels (b)
    and (c) were made with
    $\gamma_\xi/\gamma_\sigma=P_\sigma/\gamma_\sigma=1$. Panel (d)
    shows the case for the coherent excitation of \emph{all} the SPS
    by the coherent field, whereas Panel (e) shows the case for the
    excitation by only the emission of the quantum source.}
  \label{fig:scheme}
\end{figure}

\begin{figure}
  \includegraphics[width=0.95\linewidth]{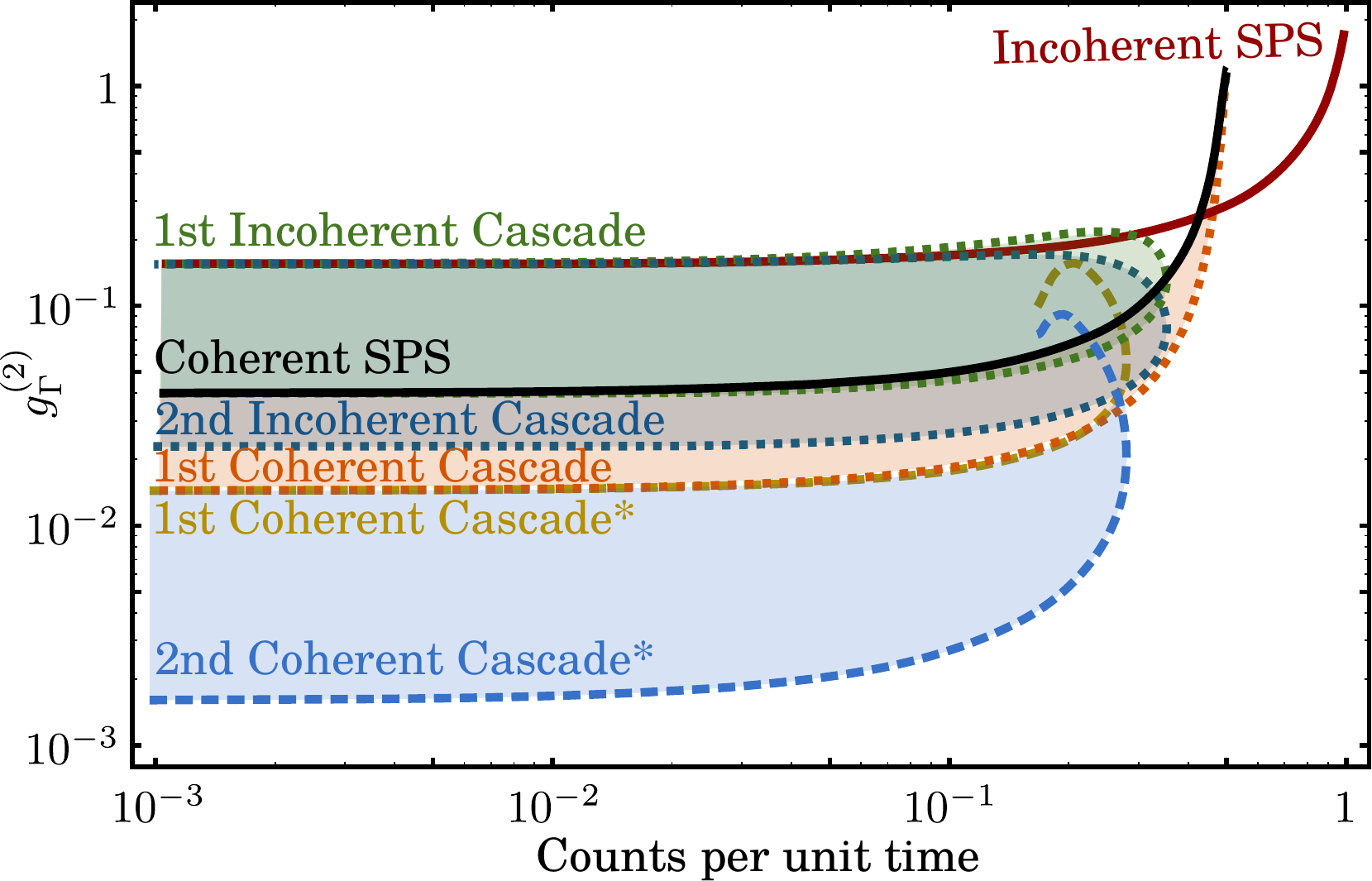}
  \caption{(Color online). Improvement in the $g^{(2)}_\Gamma$ due to
    the cascaded scheme.  Once we fixed the linewidth of the first
    SPS, the degree of freedom of the second SPS spans a new
    accessible region shown by the colour shading. The antubunching
    provided by a incoherent SPS is enhanced by the cascaded scheme,
    but for a wide ragen of parameter it is overcome by the
    antibunching provided by the coherent SPS. However, by allowing
    all the SPS in the cascade to be driven by a coherent field, a
    destructive interference sets a limits the enhancement of the
    $g^{(2)}_\Gamma$ at the first step. By driving the coherent
    cascade with only the emission of the coherent SPS (lines noted by
    ``Coherent Cascade$^{\ast}$'', but we unbound the enhancement of
    the $g^{(2)}_\Gamma$.}
  \label{fig:fig10}
\end{figure}

Here more than in the other cases discussed so far, a fully-integrated
approach might be desirable to realize a device. The scheme consists
in feeding the output field of the $i$-th system to the input field of
the $(i+1)$-th system, with no feedback.  This could be achieved by
unidirectional couplers in the laboratory or, as already commented in
Part~I\cite{arXiv_lopezcarreno16a}, by using chiral waveguides to
assemble the whole architecture compactly on the same
chip.\cite{petersen14a,coles16a} It could also be possible to use a
highy directional source of excitations.\cite{ma15b} A schematic
representation is shown in Fig.~\ref{fig:scheme}(a) and reads as
follows: the first SPS is driven by an external pumping, and the
emission of SPS1 is then sent to SP2, and so on and so forth as more
stages of the cascaded are arranged. An array of $N$ SPS with
annihilation operators $\sigma_i$, Hamiltonian $H_i$, and decay rate
$\gamma_i$ can then be described by a master equation in the
form:\cite{gardiner_book00a}
\begin{multline}
 \label{eq:marfeb16002949CET2016}
 \partial_t\rho= \sum_{j=1}^{N}\left( i[\rho, H_j] +
   \frac{\gamma_k}{2} \mathcal{L}_{c_k} \rho
   -\sqrt{\gamma_j}\left[\mathcal{E}\ud{c_j}
     -\mathcal{E^\ast}c_j,\rho\right ] \right)+{}\\
 {}+\sum_{j=2}^{N}\sum_{l=1}^{j-1}\sqrt{\gamma_j \gamma_l} \left
   \lbrace [c_l \rho, \ud{c_j}] + [c_j,\rho \ud{c_l}] \right \rbrace+
 \frac{P_{c_1}}{2}\mathcal{L}_{\ud{c_1}}\rho\, ,
\end{multline}
where $\mathcal{E}$ is the amplitude of the coherent field incident
upon SPS1. The jusification for this master equation is given in the
Appendices~\ref{sec:app1} and~\ref{sec:app2}, and we have featured
explicitly an incoherent pumping of the first SPS at the rate
$P_{c_1}$, that can be set to zero to consider the effect of the
coherent driving only.
%
%\emph{Discussion.} 
The antibunching of a two-level system is, ideally, exactly zero. An
actual experiment will detect two photons even in absence of noise and
extraneous emitters, from the SPS itself. This is due to time
uncertainty that can bring together two photons close enough in time
to exhibit photon bunching. For some fixed time unit set by the
detector, the larger the emission rate, the more likely are such
spurious coincidences. In the frequency space, this is linked to tail
events that are not detected, for instance because the detector's
bandwidth~$\Gamma$ is finite. Accordingly, while detecting all photons
at all frequencies, $\Gamma\rightarrow\infty$, produces the
ideal~$g^{(2)}_\infty(\tau=0)=0$, failure to do so results in
bunching.\cite{delvalle12a} The fastest is the source, the broader is
the spectrum and the more difficult it becomes to collect all the
photons. Crucially, the spectrum is a Lorentzian and consequently has
fat tails. This means that outliers are frequent, unlike a Normal
distribution where they are exponentially suppressed until they become
safely completely negligible. A fat-tail distribution can never safely
exclude all its outliers, regardless of the filters bandwidth. There
would therefore appear to be an intrinsic limitation between emission
rate and photon antibunching. We now show how to thwart such a
predicament by trimming the fat tails. Figure~\ref{fig:fig10} shows
the combined emission rate and antibunching~$g^{(2)}_\Gamma$ for a SPS
that is excited either incoherently (solid red) or coherently (solid
black) as pumping is varied (dummy parameter on the curve). This
figure is for a detector bandwidth~$\Gamma=4\gamma_\sigma$. Increasing
$\Gamma$, one both betters the emission rate and antibunching, towards
their ideal values of~$\gamma_\sigma$ and~$0$, respectively. However,
this is a slow convergence that requires unpractical large
filters at a disadvantage with the fat Lorentzian tails.  The exact
functions of~$\Gamma$ and other parameters are given by, in the case
of incoherent pumping at rate~$P_\sigma$:
\begin{subequations}
  \label{eq:marfeb16111007CET2016}
  \begin{align}
    I^\mathrm{inc}_\sigma&=\gamma_\sigma P_{\sigma}/\left(\gamma_{\sigma} + P_{\sigma} \right)\int_{-\Gamma/2}^{\Gamma/2} S(\omega)\, d\omega \, , \\
    g^{(2),\mathrm{inc}}_\sigma&=\frac{2}{1+3 \tan\left [\pi\left( \gamma_\sigma +
          P_\sigma \right)  I^\mathrm{inc}_\sigma/ \left( 2\gamma_\sigma P_\sigma
        \right) \right]}\, ,\label{eq:marfeb16183516CET2016}
  \end{align}
\end{subequations}
where Eq.~(\ref{eq:marfeb16183516CET2016}) is valid for
$I^\mathrm{inc}_{\sigma} \leq \gamma_\sigma P_{\sigma}/\left(\gamma_{\sigma} + P_{\sigma} \right)$.
%\subsection{Coherent driving}
The coherent counterpart is given by:
\begin{widetext}
  \begin{subequations}
    \begin{align}
      \label{eq:marfeb16121648CET2016}
      I^\mathrm{coh}_\sigma&=\frac{4\gamma_\sigma\Omega^2}{\gamma_\sigma^2+8\Omega^2}\int_{-\Gamma/2}^{\Gamma/2} S(\omega)\, d\omega \, , \\
      g^{(2),\mathrm{coh}}_\sigma&= \lbrace 2\gamma_{\widetilde{11}}(\gamma_{\widetilde{01}}^2+8\Omega^2)(\gamma_{\widetilde{11}}\gamma_{\widetilde{12}}+16\Omega^2)\times\nonumber\\
      {}\times[&\gamma_{\widetilde{11}}^2\gamma_{\widetilde{21}}\gamma_{\widetilde{31}}\gamma_{\widetilde{12}}\gamma_{\widetilde{32}}(9\gamma_{\widetilde{10}}^2+7\gamma_{\widetilde{10}}\gamma_{\widetilde{01}}+\gamma_{\widetilde{01}}^2) + 4\gamma_{\widetilde{11}}\gamma_{\widetilde{32}}(84\gamma_{\widetilde{10}}^4+16\gamma_{\widetilde{10}}^3\gamma_{\widetilde{01}}+118 \gamma_{\widetilde{10}}^2\gamma_{\widetilde{01}}^2+31 \gamma_{\widetilde{10}}\gamma_{\widetilde{01}}^3+2\gamma_{\widetilde{01}}^4)\Omega^2+{}\nonumber\\&{}+32 \gamma_{\widetilde{10}}(51\gamma_{\widetilde{10}}^3+75\gamma_{\widetilde{10}}^2\gamma_{\widetilde{01}}+38 \gamma_{\widetilde{10}}\gamma_{\widetilde{01}}^2+8\gamma_{\widetilde{01}}^3)\Omega^4+768\gamma_{\widetilde{10}}^2\Omega^6 ] \rbrace \big /  \nonumber\\
                     &\qquad\left \lbrace
                             3\gamma_{\widetilde{31}}(\gamma_{\widetilde{11}}\gamma_{\widetilde{21}}+4\Omega^2)
                             (\gamma_{\widetilde{11}}\gamma_{\widetilde{21}}+8\Omega^2)(\gamma_{\widetilde{31}}
                             \gamma_{\widetilde{32}}+16\Omega^2)(\gamma_{\widetilde{11}}^2\gamma_{\widetilde{12}}+
                             8\gamma_{\widetilde{10}}\Omega^2)^2\right
                             \rbrace\, ,
    \end{align}
  \end{subequations}
\end{widetext}
where we have used again the compact notation
$\gamma_{\widetilde{nm}}^k = (n\Gamma + m \gamma_\sigma)^k$.  From
these results, and as is apparent on Fig.~\ref{fig:fig10}, one sees
that the coherent driving provides better antibunching than its
incoherent counterpart, with plateaus of
$g^{(2),\mathrm{coh}}_\sigma=2\gamma_{\widetilde{01}}^2(9\gamma_{\widetilde{10}}^2+7\gamma_{\widetilde{01}}\gamma_{\widetilde{10}}+\gamma_{\widetilde{01}}^2)/3\gamma_{\widetilde{11}}^2\gamma_{\widetilde{21}}\gamma_{\widetilde{31}}$
and
$g^{(2),\mathrm{inc}}_\sigma=2\gamma_{\widetilde{01}}/\gamma_{\widetilde{31}}$,
which for $\Gamma=4\gamma_\sigma$ reduce to~$346/8775~\approx~0.039$
and $2/13~\approx~0.154$ respectively.  The maximum emission rate,
however, is smaller. This is due to stimulated emission induced by the
coherent driving that Rabi oscillates back the excited state to the
ground state and thus saturates the SPS to at most half its full
occupancy. In contrast, the incoherent driving can saturate the
two-level system to its excited state and can thus emit twice as much.
The reason for a better antibunching of the coherent driving has
already been discussed above in relation to the Heitler effect. It is
related to the spectral shape that trims its fat tails by destructive
interferences. In the case of the Mollow triplet, we have shown above
that it consists of three Lorentzians centered at $\omega=0$, and
$\omega=\pm\sqrt{64\Omega^2-\gamma_\sigma^2}$ as shown in
Eq.~(\ref{eq:mollow}).  In the limit of vanishing driving, the
resulting lineshape thus concentrates its emission to the central peak
with a distribution of the type:
\begin{equation}
S(\omega)=\frac{1}{\pi} \frac{32 \gamma_\sigma \Omega^2}{(\gamma_\sigma^2+\omega^2)^2}\, ,
\end{equation}
which is proportional to a Student's $t$--distribution or order 3. The
tails of this distribution vanish faster than those of a Lorentzian
distribution.  A given frequency windows collects more photons from
the SPS and its antibunching is therefore more like that of the full
spectrum, that provides the exact zero.

We can extend this principle to cascaded systems, where the driving of
the SPS is not from an external classical laser (which, since it does
not get feedback, can be seen as a particular case of cascading) but
from another 2LS. For the case of incoherent pumping, we observe a
similar behavior in the emission spectrum, as a difference of
lorentzians leads to a distribution with faster decaying
tails. Namely, the emission specrtum of the first cascade reads at
vanishing pumping:
  \begin{align}
    \label{eq:viefeb5153053CET2016}
    S_\xi (\omega)&=\frac{1}{2\pi}\frac{\gamma_\sigma
                    \gamma_\xi}{\gamma_\xi-\gamma_\sigma}\left[
                    \frac{1}{(\gamma_\sigma/2)^2+\omega^2}-\frac{1}{(\gamma_\xi/2)^2+\omega^2}
                    \right]\, ,\nonumber \\
                    &= \frac{1}{2\pi}\frac{\gamma_\sigma \gamma_\xi \left(\gamma_\sigma+\gamma_\xi\right)}{(\gamma_\sigma^2 +4\omega^2 )( \gamma_\xi^2 +4\omega^2)}\, ,
  \end{align}
which is a Student~$t$ distribution of order~3. In the next stage in
the cascade, the resulting distribution is a Student~$t$ of
order~5.

There is a trade-off between the antibunching and the intensity of the
emission of the cascaded single-photon source. The improvement of the
value of $g^{(2)}_\Gamma$ is maximum when all the SPS have the same
decay rate. In Fig.~\ref{fig:fig10} we show the antibunching as a
function of the emission rate of the cascaded SPS at various steps of
the cascade, and for both the coherent and the incoherent driving. The
coherent (solid, black line) and incoherent (solid, red line) are
simply a SPS driven by a classical field. In the first step of the
cascade we obtain a large enhancement in the antibunching: for a wide
range of intensities, the antibunching for the first step in the
incoherent cascade (dotted, green line) matches the antibunching
obtained with the coherent SPS, which shows again how driving a system
with a classical laser can be seen as cascade. On the other hand, the
antibunching for the first step in the coherent cascade (dotted,
orange line) shows a huge enhancement, which for a large range of
intensities is even better that the antibunching for the second step
in the incoherent cascade (dotted, blue line). However, the second
step in the coherent cascade shows the same antibunching as the
coherent SPS, thus effectivelly reducing its antibunching. In fact, if
all the SPS in the cascade have the same decay rate, then all the even
steps in the coherent cascade have the antibunching of the coherent
SPS, whereas the odd steps have the antibunching of the first step in
the coherent cascade. Another possible configuration for the coherent
cascade provides a gain in the $g^{(2)}$ in the second step of the
cascade. Namely, removing the coherent excitation from all the SPS
except for the first one. This is achieved in the same way as in
Section~\ref{sec:juefeb25193818CET2016}, for which
Eq.~\ref{eq:marfeb16002949CET2016} is modified to:
\begin{multline}
\label{eq:juejul21162238CET2016}
 \partial_t\rho= \sum_{j=1}^{N}\left( i[\rho, H_j] +
   \frac{\gamma_k}{2} \mathcal{L}_{c_k}\right) \rho
   -\sqrt{\epsilon \gamma_1}\left[\mathcal{E}\ud{c_1}
     -\mathcal{E^\ast}c_1,\rho\right ]+{}\\
 {}+\sum_{j=3}^{N}\sum_{l=2
}^{j-1}\sqrt{\gamma_j \gamma_l} \left
   \lbrace [c_l \rho, \ud{c_j}] + [c_j,\rho \ud{c_l}] \right \rbrace+{}\\
{}+\sum_{j=2}^{N}\sqrt{(1-\epsilon)\gamma_1 \gamma_j} \left
   \lbrace [c_1 \rho, \ud{c_j}] + [c_j,\rho \ud{c_1}] \right \rbrace\, .
\end{multline}
The coherent SPS provided by Eq.~(\ref{eq:juejul21162238CET2016}) has
the same antibunching as the one provided by
Eq.~(\ref{eq:marfeb16002949CET2016}), but requires a larger amplitude
of the coherent field $\mathcal{E}$ as it is effectively reduced by a
factor $\sqrt{\epsilon}$. For a large range of values of the emission
rate, the antibunching of the first step in the cascade coincides with
the values obtained with
Eq.~(\ref{eq:marfeb16002949CET2016}). However, only a fraction
$1-\epsilon$ of the light emitted by the coherent SPS is transmitted
to the cascade, and the emission rate of the cascade is reduced by
that factor. This is shown by the dashed, golden line in
Fig.~\ref{fig:fig10}. The antibunching for the second step in the
cascade, as opposed to what is obtained with
Eq.~(\ref{eq:marfeb16002949CET2016}), improves vastly as shown by the
dashed, blue line in Fig.~\ref{fig:fig10}.  Considering, e.g., a
typical lifetime of $\tau_\sigma=1\,\mathrm{ns}$ for a self-assembled
quantum dot, which transform-limit turns into a PL line of
$\gamma_\sigma=\hbar/\tau_\sigma=666\, \mathrm{eV}$, that is,
furthermore, power-broadened by pumping, one gets emission rates of
ten million counts per second with an antibunching of
$\approx 3.5\times10^-3$ in a frequency window of $4\gamma_\sigma$
with two cascades under incoherent pumping, and one order of magnitude
better antibunching with coherent pumping. The cascading can be
iterated further and by turning to shorter lived two-level emitters,
since we have removed the spectral constrain of outliers spoiling the
antibunching, we can reach extremely bright antibunched single photon
sources.  To appreciate these results, one can stress the saturated
antibunching for the incoherent (red) an coherent (black) uncascaded
SPS while SPSs in the cascades better their antibunching, albeit at
the cost of diminished emission rates. This is because the first SPS
sets the unit of time. In fact, the envelope of the first cascade, in
the units fixed by the first SPS, is the same as the envelope of this
first SPS .  As a consequence, what one achieves already at the first
case of the cascade is to increase the decay rate of the SPS for a
given value of the antibunching, thereby achieving brighter single
single photon sources. Next stages in the cascade further improve the
situation.

\subsection{Comparison with the excitation of an harmonic oscillator}
\label{sec:juefeb25201016CET2016}

\begin{figure}
  \includegraphics[width=0.95\linewidth]{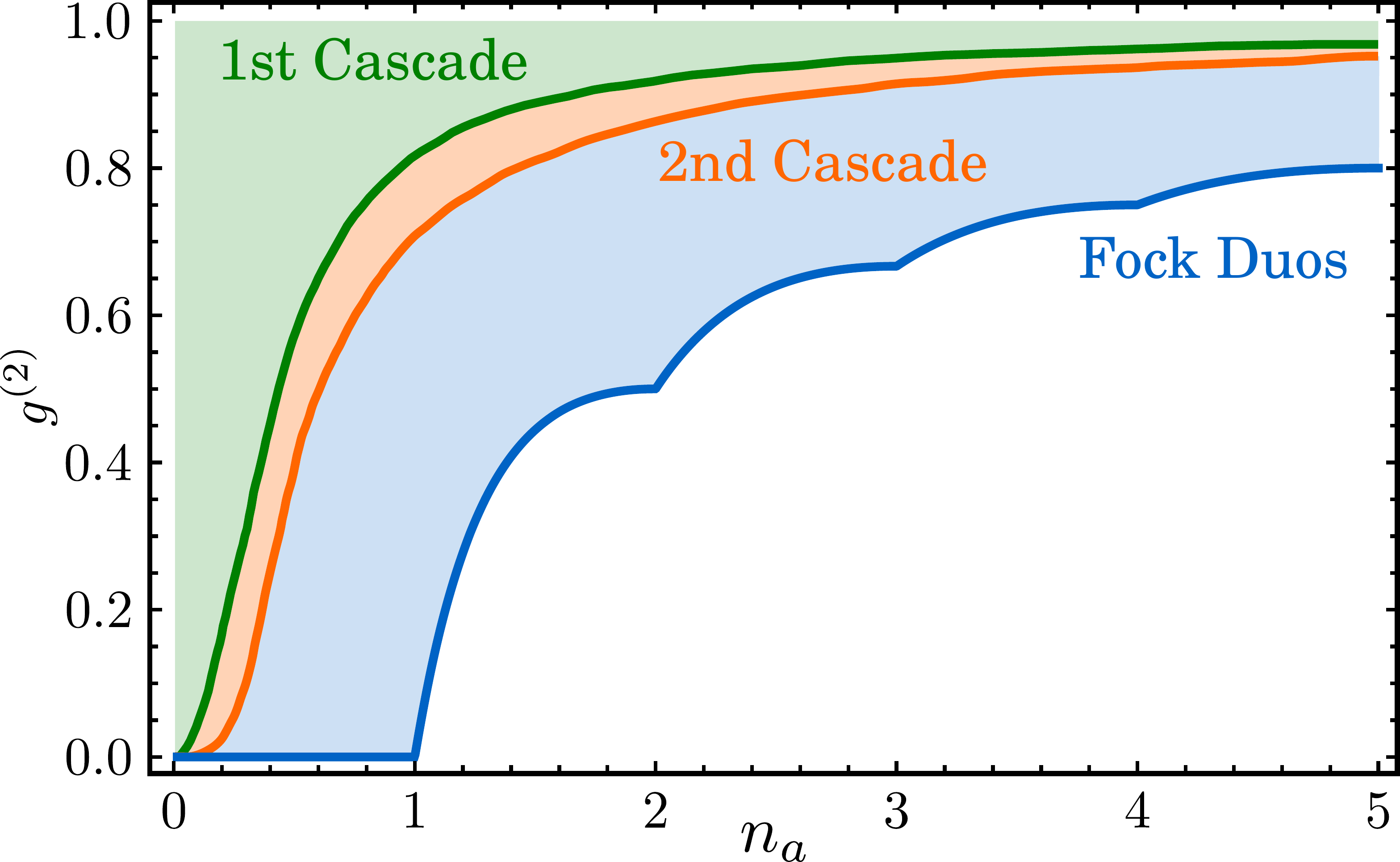}
  \caption{(Color online).  Getting closer to the limit: adding a
    cascade step to excite an harmonic oscillator pushed the border of
    the accessible states from the green to the orange region. Adding
    more steps to the cascade could take the acessible region closer
    to the Fock Duos limit.}
  \label{fig:ExcitingTheHO}
\end{figure}

In the opening Part of this series\cite{arXiv_lopezcarreno16a} we
treated the excitation of an harmonic oscillator by a quantum
source. Such excitation allowed the harmonic oscillator to cover some
area in the $(n_a,g^{(2)})$ space, shown in green in
Fig.~\ref{fig:ExcitingTheHO}(a). Since this was some distance away
from physical limit, a natural question is whether other quantum
sources could access more territory.  We now show how, indeed, the
emission from a cascaded SPS indeed achieves that goal, although still
not yet touching the boundary. In the case of driving the oscillator
with cascaded SPS, the excitation is obtained by adding
$(\gamma_a/2)\mathcal{L}_a \rho + \sqrt{\gamma_\sigma \gamma_a} \left
  \lbrace [\sigma \rho, \ud{a}] + [ a,\rho \ud{\sigma}] \right
\rbrace+ \sqrt{\gamma_\xi \gamma_a} \left \lbrace [\xi \rho, \ud{a}] +
  [ a,\rho \ud{\xi}] \right \rbrace$
to Eq.~(\ref{Eq.MasterEquation}). Here $\gamma_a$ is the decay rate,
and $a$ the annihilation operator of the harmonic oscillator.  In this
configuration, the accessible area expands to the orange region of
Fig.~\ref{fig:ExcitingTheHO}. Inspired by the positive results
obtained using a second cascade to excite the harmonic oscillator, one
might think that further cascading the source could move the border of
the accessible states closer to the limit set by the Fock duo.
However, covering all the possible values for all the free parameters
(driving intensity, decay rates and frequencies) grows exponentially
with the number of cascades in the hypothetical device.  It remains an
open question at the time of writing whether further cascades, or for
that matter another quantum source, can get closer or even reach the
frontier.

\section{Conclusions}
\label{sec:jueoct29112554CET2015}

We have studied in detail several aspects as well as several
configurations of the driving of a two-level system (2LS) by quantum
light (and classical light for comparison).

We have shown that the 2LS is ``quantum-enough'' not to benefit by
itself from the excitation by quantum light as far as the quantum
state is concerned, since classical excitation can drive it in the
same steady state. However, when granted together with its source,
with which it can enter in strong-coupling, or when considering
dynamical aspects of its emission, quantum driving of a SPS can result
in new regimes for its dynamics.  This led us to a discussion of the
meaning and definition of strong-coupling between states in absence of
feedback and, therefore, of oscillations.  

At a fundamental level, we have meticulously discussed all the main
possibilities to drive a 2PS with various types of light, yielding a
series of Mollow triplets that are summarized in
Fig.~\ref{fig:MollowDensity}. As the most elaborate description, we
have presented a full quantized model where the lasing is
self-consistently formed out of incoherent excitation through the
one-atom lasing mechanism. This allows to consider aspects such as the
question of the optical phase and the nature of the Rayleigh peak in
absence of any $c$-number in the Hamiltonian.

At an applied level, we have shown how the quality of the reduction of
multiple-photon emission (antibunching) is linked to the spectral
tails. We identified in this way why coherent pumping (resonance
fluorescence) overtakes its incoherent counterpart and discussed the
counter-intuitive Heitler effect of antibunched scattered photons off
a laser. Namely, the coherent driving leads, in the weak-coupling of
the Mollow triplet, to interferences that turn the luminescence line
into a Student-$t$ distribution with weaker fat tails than the
Lorentzian profile of spontaneous emission. This allows to collect
more easily all the signal that, when taken in its entirety, contrives
to result in a single emitted photon by destructive inteferences of
the possible emissions.  We have shown how a similar and even further
trimming of fat tails can be achieved by cascading SPS with the effect
of yielding increasingly antibunched emitters with decreasing tails,
ultimately converging towards a normal distribution (with no
outliers), making it contemplable to thus design \emph{perfect} single
photon sources, i.e., that yield exactly zero coincidence and perfect
antibunching, $g^{(2)}=0$, by considering, say, the emission in five
standard deviations of the cascaded SPS with enough stages to approach
a Gaussian. This would provide one spurious coincidence in 3.5
millions repetitions, making any departure from exactly zero to any
other factor in the experiment than the source itself.  These results
should stimulate experiments on new approaches to design better single
photon sources.

\begin{acknowledgments}
  Funding by the POLAFLOW ERC project No.~308136 and by the Spanish
  MINECO under contract FIS2015-64951-R (CLAQUE) is acknowledged.
\end{acknowledgments}
%\vfill\eject
\appendix

\section{Master equation for the cascade of two systems}
\label{sec:app1}

The theory of cascaded systems~\cite{gardiner_book00a} allows to
couple unidirectionally two quantum systems, so that one of them can
be regarded as the source of the excitation that the other one
receives, thus behaving as an optical target. The formalism is based
on writting the Langevin equation for the operators of the two systems:
\begin{subequations}
  \begin{align}
    \dot{a}_1 & = i[a_1,H_\mathrm{sys}]-[a_1,\ud{c_1}]\left \lbrace \frac{\gamma_1}{2}c_1+\sqrt{\gamma_1}b_\mathrm{in}^{(1)}\right \rbrace +{}\\
&{} +\left \lbrace \frac{\gamma_1}{2}\ud{c_1}+\gamma_1b_\mathrm{in}^{\dagger\, (1)} \right \rbrace[a_1,c_1]\, , \nonumber  \\
    \dot{a}_2 & = i[a_2,H_\mathrm{sys}]-[a_2,\ud{c_2}]\left \lbrace \frac{\gamma_2}{2}c_2+\sqrt{\gamma_2}b_\mathrm{in}^{(2)}\right \rbrace+{} \\
& {}+\left \lbrace \frac{\gamma_2}{2}\ud{c_2}+\gamma_2b_\mathrm{in}^{\dagger\, (2)} \right \rbrace[a_2,c_2]\, ,\nonumber
  \end{align}
\label{eq:Langevin1}
\end{subequations}
where $a_k$ is any operator of system~$k$ that decays at a
rate~$\gamma_k$, $H_{\mathrm{sys}}=H_1+H_2$ is the total Hamiltonian,
describing the \emph{independent} dynamics of the two systems, and
$b_\mathrm{in}^{(k)}$ is the input field of system~$k$. Since we are
considering that the system 2 (the target) is the driven by the
emission of system 1 (the source), is clear that the input field of
system 2 corresponds to the output field of system 1:
\begin{equation}
  b_\mathrm{in}^{(2)}=b_\mathrm{in}^{(1)}+\sqrt{\gamma_1}c_1\, ,
\end{equation}
where the last term of the rhs describe the emission of system 1, due
only to its internal dynamics. Converting Eq.~(\ref{eq:Langevin1}) to
a Quantum Ito equation and assuming that the input field has an
amplitude $\mathcal{E}$, we can obtain a master equation for the
operator $\rho(t)$ describing the cascaded system, which takes the form:
\begin{multline}
  \label{eq:MasterEquation1}
  \partial_t\rho=i[\rho, H_1 + H_2] + \frac{\gamma_1}{2}
  \mathcal{L}_{c_1} \rho
  + \frac{\gamma_2}{2} \mathcal{L}_{c_2} \rho -{}\\
  {}-
  \left[\mathcal{E}(\sqrt{\gamma_1}\ud{c_1}+\sqrt{\gamma_2}\ud{c_2})
  -\mathcal{E^\ast}(\sqrt{\gamma_1}c_1+\sqrt{\gamma_2}c_2),\rho\right ]+ {}\\
  {}+\sqrt{\gamma_1 \gamma_2} \left \lbrace [c_1 \rho, \ud{c_2}] +
    [c_2,\rho \ud{c_1}] \right \rbrace  \, ,
\end{multline}
which is the equation to which we refer to in
Section~\ref{sec:juefeb25193818CET2016}. We note that the second line
in Eq.~(\ref{eq:MasterEquation1}) is a Hamilonian--like term, that
describes the coherent driving of the two systems in the cascade.

\section{Master equation for the cascade of an arbitrary number of systems}
\label{sec:app2}

The formalism presented in Appendix~\ref{sec:app1} can be generalized
to the cascade of an arbitrary number of systems. In such case, each
operator of the~$k$th system would satisfy the Langevin equation,
\begin{align}
  \dot{a}_k & = i[a_k,H_\mathrm{sys}]-[a_k,\ud{c_k}]\left \lbrace \frac{\gamma_k}{2}c_k\sqrt{\gamma_k}b_\mathrm{in}^{(k)}\right \rbrace +{}\\
            &{} +\left \lbrace \frac{\gamma_k}{2}\ud{c_k}+\gamma_k b_\mathrm{in}^{\dagger\, (k)} \right \rbrace[a_k,c_k]\, ,
\end{align}
where now the input field of the~$k$th system corresponds to the
output field of the~$(k-1)$th system, i.e.,
$b_{\mathrm{in}}^{(k)}=b_{\mathrm{in}}^{(1)}+\sum_{j=1}^{k-1}\sqrt{\gamma_j}c_j$. As
in the Appendix~\ref{sec:app1}, we can derive the master equation for
operator $\rho(t)$ describing the cascaded system,
\begin{multline}
  \label{eq:MasterEquation1}
  \partial_t\rho= \sum_{j=1}^{N}\left( i[\rho, H_j] +
    \frac{\gamma_k}{2} \mathcal{L}_{c_k} \rho
  -\sqrt{\gamma_j}\left[\mathcal{E}\ud{c_j}
    -\mathcal{E^\ast}c_j,\rho\right ] \right)+{}\\
  {}+\sum_{j=2}^{N}\sum_{l=1}^{j-1}\sqrt{\gamma_j \gamma_l} \left
    \lbrace [c_l \rho, \ud{c_j}] + [c_j,\rho \ud{c_l}] \right \rbrace \, ,
\end{multline}
which is the equation that we refer to in
Section~\ref{sec:juefeb25200943CET2016}.

\bibliographystyle{naturemag}
\bibliography{Sci,arXiv,books} 

\begin{thebibliography}{10}
\expandafter\ifx\csname url\endcsname\relax
  \def\url#1{\texttt{#1}}\fi
\expandafter\ifx\csname urlprefix\endcsname\relax\def\urlprefix{URL }\fi
\providecommand{\bibinfo}[2]{#2}
\providecommand{\eprint}[2][]{\url{#2}}

\bibitem{feynman_book71c}
\bibinfo{author}{Feynman, R.~P.}, \bibinfo{author}{Leighton, R.~B.} \&
  \bibinfo{author}{Sands, M.}
\newblock \emph{\bibinfo{title}{Feynman Lectures on Physics}}, vol.
  \bibinfo{volume}{III} (\bibinfo{publisher}{Addison Wesley},
  \bibinfo{year}{1971}).

\bibitem{deutsch85a}
\bibinfo{author}{Deutsch, D.}
\newblock \bibinfo{title}{Quantum theory, the {C}hurch-{T}uring principle and
  the universal quantum computer}.
\newblock \emph{\bibinfo{journal}{Proc. R. Soc. Lond. A}}
  \textbf{\bibinfo{volume}{400}}, \bibinfo{pages}{97} (\bibinfo{year}{1985}).

\bibitem{golubev14a}
\bibinfo{author}{Golubev, N.~V.} \& \bibinfo{author}{Kuleff, A.~I.}
\newblock \bibinfo{title}{Control of populations of two-level systems by a
  single resonant laser pulse}.
\newblock \emph{\bibinfo{journal}{Phys. Rev. A}} \textbf{\bibinfo{volume}{90}},
  \bibinfo{pages}{035401} (\bibinfo{year}{2014}).

\bibitem{allen_book87a}
\bibinfo{author}{Allen, L.} \& \bibinfo{author}{Eberly, J.~H.}
\newblock \emph{\bibinfo{title}{Optical Resonance and Two-Level Atoms}}
  (\bibinfo{publisher}{Dover}, \bibinfo{year}{1987}).

\bibitem{arXiv_lopezcarreno16a}
\bibinfo{author}{{L\'opez Carre\~no}, J.~C.} \& \bibinfo{author}{Laussy, F.~P.}
\newblock \bibinfo{title}{Exciting with quantum light. i. exciting an harmonic
  oscillator}.
\newblock \emph{\bibinfo{journal}{arXiv:1601.06187}}  (\bibinfo{year}{2016}).

\bibitem{mollow69a}
\bibinfo{author}{Mollow, B.~R.}
\newblock \bibinfo{title}{Power spectrum of light scattered by two-level
  systems}.
\newblock \emph{\bibinfo{journal}{Phys. Rev.}} \textbf{\bibinfo{volume}{188}},
  \bibinfo{pages}{1969} (\bibinfo{year}{1969}).

\bibitem{bengtsson_book08a}
\bibinfo{author}{Bengtsson, I.} \& \bibinfo{author}{\.{Z}yczkowski, K.}
\newblock \emph{\bibinfo{title}{Geometry of Quantum States}}
  (\bibinfo{publisher}{Cambridge University Press}, \bibinfo{year}{2008}).

\bibitem{feynman57a}
\bibinfo{author}{Feynman, R.~P.}, \bibinfo{author}{F.~L.~Vernon, J.} \&
  \bibinfo{author}{Hellwarth, R.~W.}
\newblock \bibinfo{title}{Geometrical representation of the {Schr\"odinger}
  equation for solving {Maser} problems}.
\newblock \emph{\bibinfo{journal}{J. Appl. Phys.}}
  \textbf{\bibinfo{volume}{28}}, \bibinfo{pages}{49} (\bibinfo{year}{1957}).

\bibitem{jaynes63a}
\bibinfo{author}{Jaynes, E.} \& \bibinfo{author}{Cummings, F.}
\newblock \bibinfo{title}{Comparison of quantum and semiclassical radiation
  theory with application to the beam maser}.
\newblock \emph{\bibinfo{journal}{Proc. IEEE}} \textbf{\bibinfo{volume}{51}},
  \bibinfo{pages}{89} (\bibinfo{year}{1963}).

\bibitem{delvalle09a}
\bibinfo{author}{del Valle, E.}, \bibinfo{author}{Laussy, F.~P.} \&
  \bibinfo{author}{Tejedor, C.}
\newblock \bibinfo{title}{Luminescence spectra of quantum dots in
  microcavities. {II. Fermions}}.
\newblock \emph{\bibinfo{journal}{Phys. Rev. B}} \textbf{\bibinfo{volume}{79}},
  \bibinfo{pages}{235326} (\bibinfo{year}{2009}).

\bibitem{quesada12a}
\bibinfo{author}{Quesada, N.}
\newblock \bibinfo{title}{Strong coupling of two quantum emitters to a single
  light mode: The dissipative {Tavis}--{Cummings} ladder}.
\newblock \emph{\bibinfo{journal}{Phys. Rev. A}} \textbf{\bibinfo{volume}{86}},
  \bibinfo{pages}{013836} (\bibinfo{year}{2012}).

\bibitem{delvalle11a}
\bibinfo{author}{del Valle, E.} \& \bibinfo{author}{Laussy, F.~P.}
\newblock \bibinfo{title}{Regimes of strong light-matter coupling under
  incoherent excitation}.
\newblock \emph{\bibinfo{journal}{Phys. Rev. A}} \textbf{\bibinfo{volume}{84}},
  \bibinfo{pages}{043816} (\bibinfo{year}{2011}).

\bibitem{cohentannoudji77a}
\bibinfo{author}{Cohen-Tannoudji, C.~N.} \& \bibinfo{author}{Reynaud, S.}
\newblock \bibinfo{title}{Dressed-atom description of resonance fluorescence
  and absorption spectra of a multi-level atom in an intense laser beam}.
\newblock \emph{\bibinfo{journal}{J. Phys. B.: At. Mol. Phys.}}
  \textbf{\bibinfo{volume}{10}}, \bibinfo{pages}{345} (\bibinfo{year}{1977}).

\bibitem{zoller78a}
\bibinfo{author}{Zoller, P.}
\newblock \bibinfo{title}{Emission spectra of atoms strongly driven by finite
  bandwidth laser light}.
\newblock \emph{\bibinfo{journal}{J. Phys. B.: At. Mol. Phys.}}
  \textbf{\bibinfo{volume}{11}}, \bibinfo{pages}{805} (\bibinfo{year}{1978}).

\bibitem{zoller78b}
\bibinfo{author}{Zoller, P.}
\newblock \bibinfo{title}{Atomic relaxation and resonance fluorescence in
  intensity and phase-fluctuating laser light}.
\newblock \emph{\bibinfo{journal}{J. Phys. B.: At. Mol. Phys.}}
  \textbf{\bibinfo{volume}{11}}, \bibinfo{pages}{2825} (\bibinfo{year}{1978}).

\bibitem{zoller78c}
\bibinfo{author}{Zoller, P.} \& \bibinfo{author}{Ehlotzky, F.}
\newblock \bibinfo{title}{Resonance fluorescence in phase-frequency modulated
  laser fields}.
\newblock \emph{\bibinfo{journal}{Z. Phys. A}} \textbf{\bibinfo{volume}{285}},
  \bibinfo{pages}{245} (\bibinfo{year}{1978}).

\bibitem{zoller79a}
\bibinfo{author}{Zoller, P.}
\newblock \bibinfo{title}{Resonant multiphoton ionization by finite-bandwidth
  chaotic fields}.
\newblock \emph{\bibinfo{journal}{Phys. Rev. A}} \textbf{\bibinfo{volume}{19}},
  \bibinfo{pages}{1151} (\bibinfo{year}{1979}).

\bibitem{delvalle10d}
\bibinfo{author}{del Valle, E.} \& \bibinfo{author}{Laussy, F.~P.}
\newblock \bibinfo{title}{Mollow triplet under incoherent pumping}.
\newblock \emph{\bibinfo{journal}{Phys. Rev. Lett.}}
  \textbf{\bibinfo{volume}{105}}, \bibinfo{pages}{233601}
  (\bibinfo{year}{2010}).

\bibitem{laussy09a}
\bibinfo{author}{Laussy, F.~P.}, \bibinfo{author}{del Valle, E.} \&
  \bibinfo{author}{Tejedor, C.}
\newblock \bibinfo{title}{Luminescence spectra of quantum dots in
  microcavities. {I.} {Bosons}}.
\newblock \emph{\bibinfo{journal}{Phys. Rev. B}} \textbf{\bibinfo{volume}{79}},
  \bibinfo{pages}{235325} (\bibinfo{year}{2009}).

\bibitem{weisskopf30a}
\bibinfo{author}{Weisskopf, W.} \& \bibinfo{author}{Wigner, E.}
\newblock \bibinfo{title}{Calculation of the natural line width on the basis of
  {Dirac}'s theory of light (as translated by {J. B. Sykes})}.
\newblock \emph{\bibinfo{journal}{Zeitschrift f{\"u}r Physik}}
  \textbf{\bibinfo{volume}{63}}, \bibinfo{pages}{54} (\bibinfo{year}{1930}).

\bibitem{heitler_book44a}
\bibinfo{author}{Heitler, W.}
\newblock \emph{\bibinfo{title}{The Quantum Theory of Radiation}}
  (\bibinfo{publisher}{Oxford University Press}, \bibinfo{year}{1944}).

\bibitem{nguyen11a}
\bibinfo{author}{Nguyen, H.~S.} \emph{et~al.}
\newblock \bibinfo{title}{Ultra-coherent single photon source}.
\newblock \emph{\bibinfo{journal}{Appl. Phys. Lett.}}
  \textbf{\bibinfo{volume}{99}}, \bibinfo{pages}{261904}
  (\bibinfo{year}{2011}).

\bibitem{matthiesen12a}
\bibinfo{author}{Matthiesen, C.}, \bibinfo{author}{Vamivakas, A.~N.} \&
  \bibinfo{author}{Atat\"{u}re, M.}
\newblock \bibinfo{title}{Subnatural linewidth single photons from a quantum
  dot}.
\newblock \emph{\bibinfo{journal}{Phys. Rev. Lett.}}
  \textbf{\bibinfo{volume}{108}}, \bibinfo{pages}{093602}
  (\bibinfo{year}{2012}).

\bibitem{proux15a}
\bibinfo{author}{Proux, R.} \emph{et~al.}
\newblock \bibinfo{title}{Measuring the photon coalescence time window in the
  continuous-wave regime for resonantly driven semiconductor quantum dots}.
\newblock \emph{\bibinfo{journal}{Phys. Rev. Lett.}}
  \textbf{\bibinfo{volume}{114}}, \bibinfo{pages}{067401}
  (\bibinfo{year}{2015}).

\bibitem{arXiv_alkhuzheyri16a}
\bibinfo{author}{Al-Khuzheyri, R.} \emph{et~al.}
\newblock \bibinfo{title}{Resonance fluorescence from a telecom-wavelength
  quantum dot}.
\newblock \emph{\bibinfo{journal}{arXiv:1607.02063}}  (\bibinfo{year}{2016}).

\bibitem{gonzaleztudela13a}
\bibinfo{author}{Gonz\'alez-Tudela, A.}, \bibinfo{author}{Laussy, F.~P.},
  \bibinfo{author}{Tejedor, C.}, \bibinfo{author}{Hartmann, M.~J.} \&
  \bibinfo{author}{del Valle, E.}
\newblock \bibinfo{title}{Two-photon spectra of quantum emitters}.
\newblock \emph{\bibinfo{journal}{New J. Phys.}} \textbf{\bibinfo{volume}{15}},
  \bibinfo{pages}{033036} (\bibinfo{year}{2013}).

\bibitem{laussy12d}
\bibinfo{author}{Laussy, F.~P.}, \bibinfo{author}{del Valle, E.} \&
  \bibinfo{author}{Finley, J.}
\newblock \bibinfo{title}{Universal signatures of lasing in the strong coupling
  regime}.
\newblock \emph{\bibinfo{journal}{Proc. SPIE}} \textbf{\bibinfo{volume}{8255}},
  \bibinfo{pages}{82551G} (\bibinfo{year}{2012}).

\bibitem{mu92a}
\bibinfo{author}{Mu, Y.} \& \bibinfo{author}{Savage, C.~M.}
\newblock \bibinfo{title}{One-atom lasers}.
\newblock \emph{\bibinfo{journal}{Phys. Rev. A}} \textbf{\bibinfo{volume}{46}},
  \bibinfo{pages}{5944} (\bibinfo{year}{1992}).

\bibitem{mckeever03a}
\bibinfo{author}{McKeever, J.}, \bibinfo{author}{Boca, A.},
  \bibinfo{author}{Boozer, A.~D.}, \bibinfo{author}{Buck, J.~R.} \&
  \bibinfo{author}{Kimble, H.~J.}
\newblock \bibinfo{title}{Experimental realization of a one-atom laser in the
  regime of strong coupling}.
\newblock \emph{\bibinfo{journal}{Nature}} \textbf{\bibinfo{volume}{425}},
  \bibinfo{pages}{268} (\bibinfo{year}{2003}).

\bibitem{nomura10a}
\bibinfo{author}{Nomura, M.}, \bibinfo{author}{Kumagai, N.},
  \bibinfo{author}{Iwamoto, S.}, \bibinfo{author}{Ota, Y.} \&
  \bibinfo{author}{Arakawa, Y.}
\newblock \bibinfo{title}{Laser oscillation in a strongly coupled
  single-quantum-dot--nanocavity system}.
\newblock \emph{\bibinfo{journal}{Nat. Phys.}} \textbf{\bibinfo{volume}{6}},
  \bibinfo{pages}{279} (\bibinfo{year}{2010}).

\bibitem{arXiv_gericke16a}
\bibinfo{author}{Gericke, F.} \emph{et~al.}
\newblock \bibinfo{title}{Coexistence of lasing and strong coupling in
  quantum-dot microlasers}.
\newblock \emph{\bibinfo{journal}{arXiv:1606.05591}}  (\bibinfo{year}{2016}).

\bibitem{whitehead83a}
\bibinfo{author}{Whitehead, L.~A.}
\newblock \bibinfo{title}{Domino ``chain reaction''}.
\newblock \emph{\bibinfo{journal}{Am. J. Phys.}} \textbf{\bibinfo{volume}{51}},
  \bibinfo{pages}{182} (\bibinfo{year}{1983}).

\bibitem{lindeman42a}
\bibinfo{author}{Lindeman, R.~L.}
\newblock \bibinfo{title}{The trophic-dynamic aspect of ecology}.
\newblock \emph{\bibinfo{journal}{Ecology}} \textbf{\bibinfo{volume}{23}},
  \bibinfo{pages}{399} (\bibinfo{year}{1942}).

\bibitem{bodenstein13a}
\bibinfo{author}{Bodenstein, M.}
\newblock \bibinfo{title}{Eine theorie der photochemischen
  reaktionsgeschwindigkeiten}.
\newblock \emph{\bibinfo{journal}{Z. Phys. Chem.}}
  \textbf{\bibinfo{volume}{85}}, \bibinfo{pages}{390} (\bibinfo{year}{1913}).

\bibitem{anderson39a}
\bibinfo{author}{Anderson, H.~L.}, \bibinfo{author}{Fermi, E.} \&
  \bibinfo{author}{Szilard, L.}
\newblock \bibinfo{title}{Neutron production and absorption in uranium}.
\newblock \emph{\bibinfo{journal}{Phys. Rev.}} \textbf{\bibinfo{volume}{56}},
  \bibinfo{pages}{284} (\bibinfo{year}{1939}).

\bibitem{dicke54a}
\bibinfo{author}{Dicke, R.~H.}
\newblock \bibinfo{title}{Coherence in spontaneous radiation processes}.
\newblock \emph{\bibinfo{journal}{Phys. Rev.}} \textbf{\bibinfo{volume}{93}},
  \bibinfo{pages}{99} (\bibinfo{year}{1954}).

\bibitem{schawlow58a}
\bibinfo{author}{Schawlow, A.~L.} \& \bibinfo{author}{Townes, C.~H.}
\newblock \bibinfo{title}{Infrared and optical masers}.
\newblock \emph{\bibinfo{journal}{Phys. Rev.}} \textbf{\bibinfo{volume}{112}},
  \bibinfo{pages}{1940} (\bibinfo{year}{1958}).

\bibitem{kazarinov71a}
\bibinfo{author}{Kazarinov, R.~F.} \& \bibinfo{author}{Suris, R.}
\newblock \bibinfo{title}{Possibility of amplification of electromagnetic waves
  in a semiconductor with a superlattice}.
\newblock \emph{\bibinfo{journal}{Fizika i Tekhnika Poluprovodnikov}}
  \textbf{\bibinfo{volume}{5}}, \bibinfo{pages}{797} (\bibinfo{year}{1971}).

\bibitem{faist94a}
\bibinfo{author}{Faist, J.} \emph{et~al.}
\newblock \bibinfo{title}{Quantum cascade laser}.
\newblock \emph{\bibinfo{journal}{Science}} \textbf{\bibinfo{volume}{264}},
  \bibinfo{pages}{553} (\bibinfo{year}{1994}).

\bibitem{liew13b}
\bibinfo{author}{Liew, T. C.~H.} \emph{et~al.}
\newblock \bibinfo{title}{Proposal for a bosonic cascade laser}.
\newblock \emph{\bibinfo{journal}{Phys. Rev. Lett.}}
  \textbf{\bibinfo{volume}{110}}, \bibinfo{pages}{047402}
  (\bibinfo{year}{2013}).

\bibitem{liew16a}
\bibinfo{author}{Liew, T. C.~H.} \emph{et~al.}
\newblock \bibinfo{title}{Quantum statistics of bosonic cascades}.
\newblock \emph{\bibinfo{journal}{New J. Phys.}} \textbf{\bibinfo{volume}{18}},
  \bibinfo{pages}{023041} (\bibinfo{year}{2016}).

\bibitem{paul82a}
\bibinfo{author}{Paul, H.}
\newblock \bibinfo{title}{Photon antibunching}.
\newblock \emph{\bibinfo{journal}{Rev. Mod. Phys.}}
  \textbf{\bibinfo{volume}{54}}, \bibinfo{pages}{1061} (\bibinfo{year}{1982}).

\bibitem{aaronson11a}
\bibinfo{author}{Aaronson, S.} \& \bibinfo{author}{Arkhipov, A.}
\newblock \bibinfo{title}{The computational complexity of linear optics}.
\newblock \emph{\bibinfo{journal}{Proceedings of the 43rd Annual ACM Symposium
  on Theory of Computing}} \bibinfo{pages}{333} (\bibinfo{year}{2011}).

\bibitem{rau14a}
\bibinfo{author}{Rau, M.} \emph{et~al.}
\newblock \bibinfo{title}{Free space quantum key distribution over 500 meters
  using electrically driven quantum dot single-photon sources---a proof of
  principle experiment}.
\newblock \emph{\bibinfo{journal}{New J. Phys.}} \textbf{\bibinfo{volume}{16}},
  \bibinfo{pages}{043003} (\bibinfo{year}{2014}).

\bibitem{schlehahn15a}
\bibinfo{author}{Schlehahn, A.} \emph{et~al.}
\newblock \bibinfo{title}{Single-photon emission at a rate of 143 {MHz} from
  a deterministic quantum-dot microlens triggered by a mode-locked
  vertical-external-cavity surface-emitting laser}.
\newblock \emph{\bibinfo{journal}{Appl. Phys. Lett.}}
  \textbf{\bibinfo{volume}{107}}, \bibinfo{pages}{041105}
  (\bibinfo{year}{2015}).

\bibitem{ding16a}
\bibinfo{author}{Ding, X.} \emph{et~al.}
\newblock \bibinfo{title}{On-demand single photons with high extraction
  efficiency and near-unity indistinguishability from a resonantly driven
  quantum dot in a micropillar}.
\newblock \emph{\bibinfo{journal}{Phys. Rev. Lett.}}
  \textbf{\bibinfo{volume}{116}}, \bibinfo{pages}{020401}
  (\bibinfo{year}{2016}).

\bibitem{kuhlmann15a}
\bibinfo{author}{Kuhlmann, A.~V.} \emph{et~al.}
\newblock \bibinfo{title}{Transform-limited single photons from a single
  quantum dot}.
\newblock \emph{\bibinfo{journal}{Nat. Comm.}} \textbf{\bibinfo{volume}{6}},
  \bibinfo{pages}{8204} (\bibinfo{year}{2015}).

\bibitem{unsleber16a}
\bibinfo{author}{Unsleber, S.} \emph{et~al.}
\newblock \bibinfo{title}{Highly indistinguishable on-demand resonance
  fluorescence photons from a deterministic quantum dot micropillar device with
  74\% extraction efficiency}.
\newblock \emph{\bibinfo{journal}{Opt. Express}} \textbf{\bibinfo{volume}{24}},
  \bibinfo{pages}{8539} (\bibinfo{year}{2016}).

\bibitem{somaschi16a}
\bibinfo{author}{Somaschi, N.} \emph{et~al.}
\newblock \bibinfo{title}{Near-optimal single-photon sources in the solid
  state}.
\newblock \emph{\bibinfo{journal}{Nat. Photon.}} \textbf{\bibinfo{volume}{10}},
  \bibinfo{pages}{340} (\bibinfo{year}{2016}).

\bibitem{loredo16a}
\bibinfo{author}{Loredo, J.~C.} \emph{et~al.}
\newblock \bibinfo{title}{Scalable performance in solid-state single-photon
  sources}.
\newblock \emph{\bibinfo{journal}{Optica}} \textbf{\bibinfo{volume}{3}},
  \bibinfo{pages}{433} (\bibinfo{year}{2016}).

\bibitem{gardiner_book00a}
\bibinfo{author}{Gardiner, G.~W.} \& \bibinfo{author}{Zoller, P.}
\newblock \emph{\bibinfo{title}{Quantum Noise}}
  (\bibinfo{publisher}{Springer-Verlag, Berlin}, \bibinfo{year}{2000}),
  \bibinfo{edition}{2nd} edn.

\bibitem{petersen14a}
\bibinfo{author}{Petersen, J.}, \bibinfo{author}{Volz, J.} \&
  \bibinfo{author}{Rauschenbeutel, A.}
\newblock \bibinfo{title}{Chiral nanophotonic waveguide interface based on
  spin-orbit interaction of light}.
\newblock \emph{\bibinfo{journal}{Science}} \textbf{\bibinfo{volume}{346}},
  \bibinfo{pages}{67} (\bibinfo{year}{2014}).

\bibitem{coles16a}
\bibinfo{author}{Coles, R.~J.} \emph{et~al.}
\newblock \bibinfo{title}{Chirality of nanophotonic waveguide with embedded
  quantum emitter for unidirectional spin transfer}.
\newblock \emph{\bibinfo{journal}{Nat. Comm.}} \textbf{\bibinfo{volume}{7}},
  \bibinfo{pages}{11183} (\bibinfo{year}{2016}).

\bibitem{ma15b}
\bibinfo{author}{Ma, Y.}, \bibinfo{author}{Ballesteros, G.},
  \bibinfo{author}{Zajac, J.~M.}, \bibinfo{author}{Sun, J.} \&
  \bibinfo{author}{Gerardot, D.~B.}
\newblock \bibinfo{title}{Highly directional emission from a quantum emitter
  embedded in a hemispherical cavity}.
\newblock \emph{\bibinfo{journal}{Opt. Lett.}} \textbf{\bibinfo{volume}{40}},
  \bibinfo{pages}{2373} (\bibinfo{year}{2015}).

\bibitem{delvalle12a}
\bibinfo{author}{del Valle, E.}, \bibinfo{author}{Gonz\'alez-Tudela, A.},
  \bibinfo{author}{Laussy, F.~P.}, \bibinfo{author}{Tejedor, C.} \&
  \bibinfo{author}{Hartmann, M.~J.}
\newblock \bibinfo{title}{Theory of frequency-filtered and time-resolved
  $n$-photon correlations}.
\newblock \emph{\bibinfo{journal}{Phys. Rev. Lett.}}
  \textbf{\bibinfo{volume}{109}}, \bibinfo{pages}{183601}
  (\bibinfo{year}{2012}).

\end{thebibliography}

\end{document}